\documentclass[cmbright,fleqn,referee]{envauth}

\usepackage{adjustbox,lipsum}
\usepackage{amsmath}
\usepackage{graphicx}
\usepackage{hyperref}
\usepackage{url}
\usepackage{subcaption}
\usepackage[colorinlistoftodos]{todonotes}
\usepackage{ upgreek }
\usepackage{amsfonts}
\usepackage{caption}
\usepackage{float}
\usepackage{subcaption}
\usepackage{multirow}

\def\bSig\mathbf{\Sigma}

\received{00 Month 2012}
\revised{00 Month 2012}
\accepted{00 Month 2012}

\usepackage{natbib,upgreek}

\runninghead{P.~Das and S.~Ghosal}{Analyzing Ozone Conc. by Bayesian Spatio-temporal Quantile Regression}

\begin{document}

\title{Analyzing Ozone Concentration by Bayesian Spatio-temporal Quantile Regression}

\author{P. Das\affil{a}\corrauth\ and S. Ghosal\affil{b}}

\corraddr{P. Das, Deaprtment of Statistics, North Carolina State University, 2311 Stinson Drive, Raleigh, NC 27695, USA. E-mail: pdas@ncsu.edu}

\address{\affilnum{a} North Carolina State University, NC 27695, USA\\
\affilnum{b} North Carolina State University, NC 27695, USA}

\begin{abstract}
Ground level Ozone is one of the six common air-pollutants on which the EPA has set national air quality standards. In order to capture the spatio-temporal trend of 1-hour and 8-hour average ozone concentration in the US, we develop a method for spatio-temporal simultaneous quantile regression. Unlike existing procedures, in the proposed method, smoothing across the sites is incorporated within modeling assumptions thus allowing borrowing of information across locations, an essential step when the number of samples in each location is low. The quantile function has been assumed to be linear in time and smooth over space and at any given site is given by a convex combination of two monotone increasing functions $\xi_1$ and $\xi_2$ not depending on time. A B-spline basis expansion with increasing coefficients varying smoothly over the space is used to put a prior and a Bayesian analysis is performed. We analyze the average daily 1-hour maximum and 8-hour maximum ozone concentration level data of US and California during 2006-2015 using the proposed method. It is noted that in the last ten years, there is an overall decreasing trend in both 1-hour maximum and 8-hour maximum ozone concentration level over the most parts of the US. In California, an overall a decreasing trend of 1-hour maximum ozone level is observed while no particular overall trend has been observed in the case of 8-hour maximum ozone level.
\end{abstract}

\keywords{B-spline prior; Spatio-temporal quantile regression; Block Metropolis-Hastings; US ozone data.}
\maketitle

\section{Introduction}
In the field of analysis of weather data, greenhouse gas emission data, biological experimental data, MRI scan data, it is of interest to estimate the spatio-temporal dependence structure of the response variable at different levels such as upper and lower extremes. Traditional spatio-temporal mean regression may be inappropriate in this context. For example, if the response variable is highly skewed over the time, spatio-temporal mean regression does not represent a typical relation. \\  

Quantile regression was introduced in \cite{Koenkar1978}. Generalizations and extensions of quantile regression were proposed from both frequentist and Bayesian  perspective  (\cite{Yu2001}, \cite{Kottas2001}, \cite{Gelfand2003}, \cite{Kottas2009}). A comprehensive account of various aspects of quantile regression can be found in \cite{Koenkar2005}. The above methods were proposed for a single quantile level. The main drawback of separate quantile regression for different levels is that an estimated lower quantile curve may cross an estimated upper quantile curve violating the monotonicity of the quantiles. In addition, a single quantile regression does not allow for joint inference. \\

Methods addressing the monotonicity issue in estimating multiple quantile regression have also been proposed (\cite{He1997}, \cite{Neocleous2008}, \cite{Takeuchi2004}, \cite{Takeuchi2006}, \cite{Vapnik1995}, \cite{Shim2010}, \cite{Shim2009}, \cite{Bondell2010}, \cite{Dunson2005}). The non-crossing quantile regression method proposed in \cite{Wu2009} sequentially updates the quantile curves under the constraint that the higher estimated quantile curves stay above the the lower ones. The problem with this method and the many of the above mentioned non-crossing quantile regression methods is that the estimated values of the quantile curves are dependent on the grid of quantiles where the curves are estimated. For example, using most of the above-mentioned methods, if we estimate the quantile lines for quantile grids $T_1 = \{0.25,0.5,0.75\}$ and $T_2 = \{0.2,0.5,0.8\}$, the estimates of $\tau=0.5$ by the two methods would come different, which is not desirable. \\

Instead of fitting the quantile regression lines for a fixed set of quantiles, a more informative picture emerges by estimating the entire quantile curve (\cite{Tokdar2012}, \cite{Das2016}). Let $Q(\tau|x) = \text{inf} \{q : P(Y \leq q |X=x) \geq \tau\}$ denote the $\tau$-th quantile  $(0\leq \tau\leq 1)$ of a response $Y$ at $X=x$, $X$ being the predictor. From Theorem 1 of  \cite{Tokdar2012} it follows that a linear specification $Q(\tau|x)=\beta_0(\tau) + x\beta_1(\tau), \; \tau \in [0,1]$ is monotonically increasing in $\tau$ for every $x\in[0,1]$ if and only if 
\begin{align*}
Q(\tau|x) = \mu+\gamma x+\sigma_1x\xi_1(\tau)+\sigma_2(1-x)\xi_2(\tau),
\end{align*}
where $\sigma_1$ and $\sigma_2$ are some constants and $\xi_1,\xi_2 : ][0,1] \mapsto [0,1]$ are monotonically increasing function in $\tau \in [0,1]$. A monotonic transformation of the explanatory and the response variables to unit intervals reduces to this
\begin{align}
Q(\tau|x)=x\xi_1(\tau) + (1-x)\xi_2(\tau),
\label{eq:single_quantile}
\end{align}
where $\xi_1$ and $\xi_2$ are monotonically increasing bijection from the unit interval to itself. \cite{Tokdar2012} took transformed Gaussian process prior to estimate $\xi_1$ and $\xi_2$, while \cite{Das2016} used B-spline basis for that purpose and argued the advantages of using B-spline basis over transformed Gaussian process prior.  \\

Consider the scenario where time is the explanatory variable and the values of the response variables corresponding to time-points are available for different spatial locations. The methods of estimating a single quantile regression curve has been extended to analyze spatial data (\cite{Lee2010}, \cite{Oh2011}, \cite{Sobotka2012}). Various methods for modeling spatially varying distribution function have been proposed in last decade (\cite{Dunson2008}, \cite{Gelfand2005}, \cite{Reich2007}, \cite{Griffin2006}). As mentioned in \cite{Reich2012}, these methods treat the conditional distribution of the response at each spatial location as unknown quantities and are estimated from the data. \\ 

One of the first work addressing the non-crossing issue of the spatial quantile regression  was proposed in \cite{Reich2011}. They proposed a two-stage method and ensured monotonicity of the estimated quantile functions using the Bernstein polynomial basis function. \cite{Reich2012} assumed separate non-crossing quantile functions for each spatial location which evolves over time. Spatial smoothing was performed using Gaussian process priors. Separate analysis for each spatial location may be sensitive to small sample sizes at individual locations. While this method allows estimation at all quantile levels, post-estimation processing step makes the quantification of uncertainty difficult. Instead of estimating the quantile function for a set of grid-points, the estimation of the entire quantile function is more informative. To address the small area estimation problem, instead of assuming unrelated quantile functions for each spatial locations, we assume that the quantile function is smooth over space and linear in time. For any given spatial location, the estimated quantile function has the same form as in Equation (\ref{eq:single_quantile}). Similar to \cite{Das2016}, the monotonicity constraint on the curves $\xi_1$ and $\xi_2$ are obtained through B-spline basis expansion with coefficients (which are functions of the spatial location) increasing and lying in the unit interval. The coefficients of the B-spline basis expansion over the $d$-dimensional space are modeled by the tensor product of univariate B-splines basis functions. A closed form expression for the likelihood function is obtained in terms of the parameters in $d$ independent simplexes whose dimensions depend of the degree of the used B-spline basis functions. %For estimating the coefficients of the B-spline basis functions, we use Greedy Coordinate Descent of Varying Step sizes on Multiple Simplexes (GCDVSMS) method (Das (2016)). In this algorithm, during each iteration, the objective function value is evaluated at a few feasible points (which are found using a special type of co-ordinate descent) around the current solution and the site with most favorable objective function value is taken to be the current solution for next iteration. 
Unlike the methods proposed in \cite{Reich2011} and \cite{Reich2012}, the main advantage of this method is that once we estimate the parameters of the model, estimated response variable can be found for any spatial location, time and quantiles without further kriging or interpolation, and hence allows proper uncertainty quantification within the Bayesian paradigm. Most importantly, the approach incorporates small area estimation issues automatically in its modeling approach.

\section{Ozone in troposphere}
 Ozone can be naturally found in the troposphere and other parts of the atmosphere. While, more ozone concentration in the stratosphere is desirable as it helps absorb the ultra-violet radiation, higher ozone concentration level in the troposphere is harmful for living beings. Ozone occurs naturally in the troposphere in low concentration. There are mainly two sources of tropospheric ozone. A significant amount of ozone is released by plants and soil. Other than that, sometimes ozone migrates down to the troposphere from the stratosphere. However, the extent of ground-level naturally occurring ozone concentration is not considered a threat to living beings and the environment. \\
 
On the other hand, ozone is a byproduct of many human activities. Increasing automobiles and industries are the main source the ``bad'' ozone at the ground-level. As mentioned in an online article\footnote{Source \url{http://www.windows2universe.org/earth/Atmosphere/ozone_tropo.html&edu=high}}, current ground-level ozone concentration has doubled since 1900. Although no single source emits ozone directly, it is generated when the hydrocarbons and nitrogen-oxides, emitted by automobiles, fuel power plants and other industrial machineries, interact with each other in the presence of sunlight, specifically the ultra-violet (UV) ray. In general, ozone level reaches its highest level during the summer season. Typically, in the span of a day, ozone level reaches its highest level during mid and late afternoon.\\
 
 Recently tropospheric ozone has been related to many health problems and environmental issues. Exposure to higher level of ozone might cause respiratory problems. In plants, it slows down growth and photosynthesis and damages internal cells. High ground-level ozone also damages textile dyes, rubbers and fibers and a few genre of paints. \\
 
 Because of its unstable nature, there is no existing way to move the tropospheric ozone to the stratosphere which can be thought as the ideal way to deal with this environmental hazard. Under the Clean Air act, the US Environment Protection Agency (EPA) considers tropospheric ozone as one of the six pollutants considered to be harmful to human health and environment. Proposed ozone standards can be found in this site\footnote{Source \url{https://www.epa.gov/sites/production/files/2016-04/documents/20151001basicsfs.pdf} (accessed 11-26-2016)} which says an area would meet the ozone standards if the 4th highest maximum daily 8-hour ozone concentration each year, averaged over three years, is 70 ppb (parts per billion) or below. In this paper, we develop a spatio-temporal quantile regression model for measuring the ozone concentration level over the US and California during the period 2006-2015.\\
 
Let $Q(\tau|x,\mathbf{z})$ denote the $\tau$-th quantile of the response variable i.e., ozone concentration level at location $\mathbf{z}$ and time-point $x$. We assume a linear dependence structure of the dependent variable on time and the quantile function is smooth over the space. In our model, the quantile function is given by
 \begin{align}
Q(\tau|x,\mathbf{z}) = \beta_0(\tau,\mathbf{z}) + x\beta_1(\tau,\mathbf{z}) \; \text{for} \; \tau \in [0,1].
\end{align}
 
 \section{Proposed Bayesian Method}
 \label{model_assumptions}
 Our explanatory variable $X$ is time. Let $\{X_{li}\}_{i=1}^{n_l}$ and $\{Y_{li}\}_{i=1}^{n_l}$ denote the values of $X$ and $Y$ at location $\mathbf{z}_l=(z_{1l}, \ldots,z_{dl})$ for $l=1,\ldots, L$, and let $n_l$ denote the sample size at location $\mathbf{z}_l$. By a monotonic transformation, $X$, $Y$ and each coordinates of the spatial locations are transformed to take values in the unit intervals. From Theorem 1 of \cite{Tokdar2012}, it follows that a linear specification $Q(\tau|x) = \beta_0(\tau) + x\beta_1(\tau), \; \tau \in [0,1]$ is monotonically increasing in $\tau$ for every $x \in [0,1]$ if and only if 
 \begin{align}
 Q(\tau|x) = x\xi_1(\tau) + (1-x)\xi_2(\tau) \; \text{for} \; \tau \in [0,1], \; x \in [0,1]
 \label{eq:no_space_quantile}
 \end{align}
 where $\xi_1,\xi_2:[0,1] \mapsto [0,1]$ are monotonically increasing in $\tau \in [0,1]$. Let $\mathbf{Z} = (z_1, \ldots, z_d), z_i \in [0,1]$ for $i=1,\ldots,d$, denote the spatial location (site) where the response variable is measured. We assume that the dependence structure of the quantile function at any site $\mathbf{Z} = \mathbf{z}$ is given by 
\begin{align}
Q(\tau|x,\mathbf{z}) = x\xi_1(\tau,\mathbf{z}) + (1-x)\xi_2(\tau,\mathbf{z}), \; \tau \in [0,1], \; x,y \in [0,1], \; \mathbf{z} \in [0,1]^d,
 \label{eq:space_quantile}
\end{align}
where $\xi_1(\cdot,\mathbf{z}),\xi_2(\cdot,\mathbf{z})$ are monotonic functions from $[0,1]$ to $[0,1]$. For any fixed $\mathbf{z}$, equation (\ref{eq:space_quantile}) can be reframed as
\begin{align}
Q(\tau|x,\mathbf{z}) = \beta_0(\tau,\mathbf{z}) + x\beta_1(\tau,\mathbf{z}), \; \tau \in [0,1], \; x,y \in [0,1],
\end{align}
where $\beta_0(\tau,\mathbf{z}) = \xi_2(\tau,\mathbf{z})$ and $\beta_1(\tau,\mathbf{z}) = \xi_1(\tau,\mathbf{z})-\xi_2(\tau,\mathbf{z})$ denote the slope and the intercept of type quantile regression. \\

A B-spline basis expansion is one of the most convenient approaches for estimating a function on bounded interval. To estimate $\xi_1 (\cdot,\mathbf{z})$ and $\xi_2(\cdot,\mathbf{z})$ in Equation (\ref{eq:space_quantile}), we use a B-spline basis expansion. Taking the coefficients of the B-splines basis functions in increasing order we ensure the monotonicity of these two above-mentioned functions and any monotone increasing function can be approximated through a monotone linear combination of B-splines (\cite{Boor2001}). \cite{Das2016} argued the advantages of using B-spline over Gaussian process for estimating the quantile functions. More specifically, using B-splines allows their equation (\ref{eq:why_quadratic}) to be solved analytically, significantly reducing the cost of likelihood evaluation compared with \cite{Tokdar2012}. %In case, we use quadratic B-spline (i.e., B-spline of degree 2), Equation (\ref{eq:solve}) can be solved analytically which makes the computation more efficient and accurate. Although using cubic B-spline also Equation (\ref{eq:solve}) can be solved analytically, but it increases the computational complexity. Besides, in case of quantile regression at a single site, no noticeable improvement was noted in the simulation studies in terms of estimation performance (Das and Ghosal (2016)). To reduce the computational complexity, we use quadratic B-spline for estimation purpose.
We use quadratic B-spline because then  equation (\ref{eq:why_quadratic}), which is crucial for the likelihood evaluation, reduces to a quadratic equation.\\

\subsection{Prior}
Let $0 = t_0 < t_1 < \cdots <t_{p_1} = 1$ be the equidistant knots on the interval $[0,1]$ such that $t_i-t_{i-1} = 1/p_1$ for $i=1,\ldots, p_1$. For B-spline of degree $m_1$, the number of basis functions is $J_1=p_1+m_1$. Let $\{B_{j,m_1}(\cdot)\}_{j=1}^{p_1+m_1}$ be the basis functions of B-splines of degree $m_1$ on $[0,1]$ on the above-mentioned equidistant knots. The basis expansion of the quantile functions at site $\mathbf{z}$ is given by the relations
\begin{align}
\xi_1(\tau,\mathbf{z}) &= \sum_{j=1}^{p_1+m_1}\theta_j(\mathbf{z})B_{j,m_1}(\tau), \; 0 = \theta_1(\mathbf{z}) < \cdots < \theta_{p_1+m_1}(\mathbf{z})  = 1, \nonumber \\
\xi_2(\tau,\mathbf{z}) &= \sum_{j=1}^{p_1+m_1}\phi_j(\mathbf{z})B_{j,m_1}(\tau), \; 0 = \phi_1(\mathbf{z})  < \cdots < \phi_{p_1+m_1}(\mathbf{z})  = 1,
\label{eq:constraint}
\end{align}
where the coefficients $\{\theta_j(\mathbf{z})\}_{j=1}^{p_1+m_1}$ and $\{\phi_j(\mathbf{z})\}_{j=1}^{p_1+m_1}$ are dependent on the spatial location of the data-points. To put priors on $\{\theta_j(\mathbf{z})\}_{j=1}^{k_1+m_1}$ and $\{\phi_j(\mathbf{z})\}_{j=1}^{k_1+m_1}$, we use $d$-fold tensor product of B-spline basis functions. Thus for the spatial location $\mathbf{z} = (z_1,\ldots, z_d)$, the basis expansion is given by
\begin{align}
\theta_j(\mathbf{z}) & = \sum_{k_1=1}^{p_2+m_2} \ldots\sum_{k_d=1}^{p_2+m_2} \alpha_{j k_1 \cdots k_d}B_{k_1,m_2}(z_1)\cdots B_{k_d,m_2}(z_d), \; j = 1,\ldots, p_2+m_2, \nonumber \\
\phi_j(\mathbf{z}) & = \sum_{k_1=1}^{p_2+m_2} \ldots\sum_{k_d=1}^{p_2+m_2} \beta_{j k_1 \cdots k_d}B_{k_1,m_2}(z_1)\cdots B_{k_d,m_2}(z_d) , \; j = 1,\ldots, p_2+m_2,  \nonumber \\
\end{align}
where $\{B_{k,m_2}(\cdot)\}_{k=1}^{p_2+m_2}$ are B-spline basis functions of degree $m_2$ on $[0,1]$ with equidistant knots $0=s_1 < \cdots < s_{p_2} = 1$ and $s_i-s_{i-1} = 1/p_2$ for $i=1,\ldots, p_2$. To ensure the constraints of $\{\theta_j(\mathbf{z})\}_{j=1}^{p_1+m_1}$ and $\{\phi_j(\mathbf{z})\}_{j=1}^{p_1+m_1}$ mentioned in Equation (\ref{eq:constraint}), it is sufficient to ensure the following constraints
\begin{align}
\label{parameters}
0 = \alpha_{1k_1\cdots k_d} < \cdots < \alpha_{(p_1+m_1)k_1\cdots k_d} = 1, \;
0 = \beta_{1k_1\cdots k_d} < \cdots < \beta_{(p_1+m_1)k_1\cdots k_d} = 1, 
\end{align}
for $\{k_1,\ldots, k_d\} \in \{1,\ldots, (p_2+m_2)\}^d$. \\

 Suppose that we have data for $L$ spatial sites $\mathbf{z}_1,\ldots, \mathbf{z}_L$ and at the $l$-th site, $\{X_{li},Y_{li}\}_{i=1}^{n_l}$ denote the values of the explanatory variable (time) and the response variable, $n_l$ being the number of data-points for $l=1,\ldots,L$, the log-likelihood is given by (see Appendix for the derivation)
\begin{align}
& -\sum_{l=1}^L\sum_{i=1}^{n_l}\log \bigg\{ X_{li} \sum_{j=2}^{p_1+m_1}\theta^*_j(\mathbf{z}_l)B_{j-1,m_1-1}(\tau_{X_{li}}(Y_{li},\mathbf{z}_l))   \nonumber \\
& + (1-X_{li})\sum_{j=2}^{p_1+m_1}\phi^*_j(\mathbf{z}_l)B_{j-1,m_1-1}(\tau_{X_{li}}(Y_{li},\mathbf{z}_l)) \bigg\},
\label{eq:likelihood}
\end{align}
where 
\begin{align*}
\theta^*_j(\mathbf{z}) = (p_1+m_1)(\theta_j(\mathbf{z})-\theta_{j-1}(\mathbf{z})), \;  \phi^*_j(\mathbf{z}) = (p_1+m_1)(\phi_j(\mathbf{z})-\phi_{j-1}(\mathbf{z})),
\end{align*}
for $j = 2,\ldots,p_1+m_1$. and $\tau_{X_{li}}(Y_{li},\mathbf{z}_l)$ is the solution of
\begin{align}
\label{eq:why_quadratic}
Y_{li} = X_{li}\xi_1(\tau,\mathbf{z}_l) + (1-X_{li})\xi_2(\tau,\mathbf{z}_l).
\end{align} \\

The parameters of the log-likelihood are given by Equation (\ref{parameters}). We note that once we fix the values of $k_1,\ldots, k_d$, the vector of spacings of the coefficients $\alpha_{1k_1\cdots k_d},$ $ \ldots,$ $ \alpha_{(p_1+m_1)k_1\cdots k_d}$ lie on the unit simplex. The same is true for $\beta_{1k_1\cdots k_d},$ $ \ldots, $ $ \beta_{(p_1+m_1)k_1\cdots k_d}$. Define
\begin{align}
\gamma_{jk_1\cdots k_d} = \alpha_{(j+1)k_1\cdots k_d} - \alpha_{jk_1\cdots k_d}, \; 
\delta_{jk_1\cdots k_d} = \beta_{(j+1)k_1\cdots k_d} - \beta_{jk_1\cdots k_d},
\end{align}
for $j=1,\ldots,p_1+m_1-1$ and $\{k_1,\ldots, k_d\} \in \{1,\ldots, (p_2+m_2)\}^d$. Hence for each combination of $\{k_1,\ldots, k_d\} \in \{1,\ldots, (p_2+m_2)\}^d$, we have
\begin{align}
\sum_{j=1}^{p_1+m_1-1}\gamma_{jk_1\cdots k_d} = 1 \; \text{and} \; \gamma_{jk_1\cdots k_d} \geq 0, \;
\sum_{j=1}^{p_1+m_1-1}\delta_{jk_1\cdots k_d} = 1 \; \text{and} \; \delta_{jk_1\cdots k_d} \geq 0.
\end{align}
Thus we note that the parameters of the log-likelihood can be divided into $2 \times (p_2+m_2)^d$ unit simplex blocks. We consider uniform Dirichlet prior on the unit simplex blocks $\{\gamma_{jk_1\cdots k_d}\}_{j=1}^{p_1+m_1-1}$ and $\{\delta_{jk_1\cdots k_d}\}_{j=1}^{p_1+m_1-1}$ for $\{k_1,\ldots, k_d\} \in \{1,\ldots, (p_2+m_2)\}^d$. As mentioned in the earlier section, $m_1$ is taken to be 2. The degree of the basis functions corresponding to each spatial coordinates has been considered to be cubic (i.e., $m_2=3$).

\section{Simulation study}
\label{sim_study}
In this section we compare the estimation performance estimation of the proposed method, i.e., Spline Spatio-temporal Quantile Regression (SSTQR) with piece-wise Gaussian Basis function Spatio-temporal Quantile Regression (GBSTQR) (\cite{Reich2012}) based on simulation. For the proposed SSTQR method, we consider both likelihood based and Bayesian approaches. Finding the maximum likelihood estimate (MLE) can be seen as an optimization problem of maximizing an objective function of a constrained parameter space which is given by a collection of unit simplex blocks. Since the evaluation of the likelihood involves integration and linear search, it is hard to check whether the likelihood function is convex or not. Instead of using convex optimization algorithms, global or non-convex optimization algorithms are more reasonable in this scenario. \cite{DAS_1_2016} proposed an efficient global optimization technique on a hyper-rectangular parameter space which has been shown to work faster and than existing global optimization techniques, namely Genetic Algorithm (\cite{Fraser1957}, \cite{Bethke1980}, \cite{Goldberg1989}) and Simulated annealing (\cite{Kirkpatrick1983}, \cite{Granville1994}) yielding better solutions.  \cite{DAS_2_2016} modified that algorithm for the case where the sample space is given by an unit simplex. Following that paper, \cite{DAS_3_2016} proposed `Greedy Coordinate Descent of Varying Step sizes on Multiple Simplexes' (GCDVSMS) algorithm which efficiently minimizes any non-convex (or, maximizes any non-concave) objective function of parameters given by a collection of unit simplex blocks. The main idea of GCDVSMS algorithm is making jumps of varying step-sizes within each unit simplex blocks parallelly and searching for the most favorable direction of movement. In this algorithm, every time a local solution is found, coordinate-wise jumps of various step-sizes are performed until a better solution is found. To find the MLE in this case, GCDVSMS algorithm is used. The values of the tuning parameters in GCDVSMS algorithm are chosen as follows; \textit{initial global step size} $s_{\text{initial}} = 1$, \textit{step decay rate} for the first \textit{run} $\rho_1=2$, \textit{step decay rate} for other \textit{runs} $\rho_2 = 1.5$, \textit{step size threshold} $\phi = 10^{-1}$, \textit{sparsity threshold} $\lambda=10^{-2}$, the convergence criteria controlling parameters \textit{tol\_fun\_1}=\textit{tol\_fun\_2}$=10^{-1}$,  maximum number of iterations inside each \textit{run} $\textit{max\_iter} = 5000$, maximum number of allowed \textit{runs} $\textit{max\_runs}=200$.\\

We consider the spatial locations are uniformly distributed over the unit square i.e., $[0,1]^2$. Number of spatial locations is taken to be $L=50$. For simulation purpose, three different numbers of equidistant temporal data-points for each site have been considered which are $n = 5,10,20$ (note that, $n_l=n$ for $l=1,\ldots,50$). Let $\{\mathbf{z}_l\}_{l=1}^L$ denote the spatial locations of the available sites of data where $\mathbf{z}_l=(z_{l1},z_{l2})$ is a two-tuple such that $0 \leq z_{l1}, z_{l2} \leq 1$ for $l=1,\ldots, L$. Consider the spatial quantile function $Q(\tau|x,\mathbf{z}) = x\xi_1(\tau,\mathbf{z}) + (1-x)\xi_2(\tau,\mathbf{z})$ where
\begin{align}
\xi_1(\tau,\mathbf{z}) = & (1-\frac{\mathbf{z}_{l1}+\mathbf{z}_{l2}}{2})\tau^2+\mathbf{z}_{l1}\frac{\log(1+\tau)}{2\log(2)}+\frac{\mathbf{z}_{l2}}{2}\tau^3,  \nonumber \\
\xi_2(\tau,\mathbf{z}) = & (1-\mathbf{z}_{l2}^2)\sin(\frac{\pi\tau}{2})+\mathbf{z}_{l2}^2\frac{(e^\tau-1)}{(e-1)}.
\end{align}
Note that for any given $\mathbf{z} \in [0,1]^2$, $\xi_1(\cdot,\mathbf{z})$ and $\xi_1(\cdot,\mathbf{z})$ are strictly increasing function from $[0,1]$ to $[0,1]$ satisfying $\xi_1(0,\mathbf{z})=\xi_2(0,\mathbf{z})=0$ and $\xi_1(1,\mathbf{z})=\xi_2(1,\mathbf{z})=1$. Since the quantile function is the inverse of the cumulative distribution function, taking $U \sim U(0,1)$, $Q(U|x,\mathbf{z})$ has conditional quantile function $Q(\tau|x,\mathbf{z})$. For each locations $\{\mathbf{z}_l\}_{l=1}^L$, we considered the same set of equidistant time-points such that $0=x_{l1} < \cdots <x_{ln}=1$ and $(x_{li} - x_{l(i-1)}) = 1/(n-1)$ for $i=2,\ldots, n$. We simulate the response variable $Y$ using the equation
\begin{align*}
y_{li} = x_{li}\xi_1(U_{li},\mathbf{z}_l) + (1-x_{li})\xi_2(U_{li},\mathbf{z}_l), \; l = 1,\ldots,L, \; i = 1,\ldots,n.
\end{align*}
where $U_{li} \sim U(0,1)$ for $l = 1,\ldots,L, \; i = 1,\ldots,n$. Using both methods, for each of the  locations with the number of data-points $n=5,10,20$, the quantile curves are estimated for the quantiles $\{\tau_t\}_{t=1}^{T}$ where $\tau_t = 0.05t$ and $T=19$. The above-mentioned simulation study is repeated $S=50$ times under different random number generating seeds.  Let at the $s$-th simulation study with fixed number of data-points ($n=5,10, 20$) at each site, the estimated intercept and slope at $\tau_t$-th quantile and location $\mathbf{z}_l$ are $\hat{\beta}_0^{(s)}(\tau_{t},\mathbf{z}_{l})$ and $\hat{\beta}_1^{(s)}(\tau_{t},\mathbf{z}_{l})$ respectively for $s=1,\ldots,S$, $t=1,\ldots,T$ and $l=1,\ldots, L$. $Q(\tau_{t} | x,\mathbf{z}_{l})$ denotes the true value of the $\tau_t$-th quantile at location $\mathbf{z}_l$ at time-point $X=x$ and $\hat{Q}_1^{(s)}(\tau_{t}|x,\mathbf{z}_{l})$ is the estimated value of it based on $s$-th simulation study. Then the average of mean squared error (MSE) of the slope, intercept and the quantile value at a given time-point $X=x$ are given by
\begin{align*}
T_1 = & \frac{1}{SLT}\sum_{s=1}^{S}\sum_{l=1}^{L}\sum_{t=1}^T(\beta_0(\tau_{t},\mathbf{z}_{l})-\hat{\beta}_0^{(s)}(\tau_{t},\mathbf{z}_{l}))^2 \\
T_2 = & \frac{1}{SLT}\sum_{s=1}^{S}\sum_{l=1}^{L}\sum_{t=1}^T(\beta_1(\tau_{t},\mathbf{z}_{l})-\hat{\beta}_1^{(s)}(\tau_{t},\mathbf{z}_{l}))^2 \\
T_x = & \frac{1}{SLT}\sum_{s=1}^{S}\sum_{l=1}^{L}\sum_{t=1}^T(Q(\tau_{t} | x,\mathbf{z}_{l})-\hat{Q}^{(s)}(\tau_{t}|x,\mathbf{z}_{l}))^2.
\end{align*}

\begin{table}
\centering
\resizebox{\columnwidth}{!}{%
\bgroup
\def\arraystretch{1.2}%
\begin{tabular}{|c|c|c|c|c|c|c|}
\hline
\multirow{2}{*}{\begin{tabular}[c]{@{}c@{}}Sample size \\ (per site)\end{tabular}} & \multirow{2}{*}{Methods} & \multicolumn{5}{c|}{MSE} \\ \cline{3-7} 
 &  & $\beta_0(\tau,\mathbf{z})$ & $\beta_1(\tau,\mathbf{z})$ & $Q(\tau|0.2,\mathbf{z})$ & $Q(\tau|0.5,\mathbf{z})$ & $Q(\tau|0.8,\mathbf{z})$ \\ \hline
\multirow{3}{*}{$n=5$} & SSTQR (ML) & 0.0265 & 0.0437 & 0.0194 & 0.0154 & 0.0192 \\ \cline{2-7} 
& SSTQR (Bayes) & 0.0410 & 0.0162 & 0.0103 & 0.0076 & 0.0122 \\ \cline{2-7} 
 & GBSTQR & 0.2014 & 0.0349 & 0.0471 & 0.1159 & 0.2209 \\ \hline
\multirow{3}{*}{$n=10$} & SSTQR (ML) & 0.0248 & 0.0340 & 0.0187 & 0.0146 & 0.0166 \\ \cline{2-7} 
& SSTQR (Bayes) & 0.0290 & 0.0127 & 0.0081 & 0.0055 & 0.0082 \\ \cline{2-7} 
 & GBSTQR & 0.6000 & 0.0241 & 0.0658 & 0.2337 & 0.5097 \\ \hline
\multirow{3}{*}{$n=20$} & QSSTQR (ML) & 0.0186 & 0.0239 & 0.0145 & 0.0119 & 0.0136 \\ \cline{2-7} 
& SSTQR (Bayes) & 0.0154 & 0.0100 & 0.0072 & 0.0052 & 0.0061 \\ \cline{2-7} 
 & GBSTQR & 0.0685 & 0.0171 & 0.0165 & 0.0398 & 0.0754 \\ \hline
\end{tabular}
\egroup
}
\caption{Average MSE (based on $S = 50$ repetitions of simulation study) of estimated slope, intercept and estimated quantile function value at three time-points $X=0.2, 0.5, 0.8$ using SSTQR (both MLE based and Bayesian approach) and GBSTQR over all $L=50$ locations and $T=19$ quantile levels $\tau = 0.05, 0.10, \ldots, 0.95$.}
\label{mse_table}
\end{table}

For SSTQR (Bayes) method, 10000 iterations are performed and the first 1000 iterations are disregarded as burn-in. In the Table (\ref{mse_table}), the comparison of average MSE of estimated slope, intercept and estimated quantile curve value at three time-points $X=0.2, 0.5, 0.8$ over all locations are provided for both of the above-mentioned methods. Since we are using quadratic B-spline for the quantile basis functions and cubic B-spline for the coordinate-wise spatial basis functions, $m_1= 2$, $m_2=3$. To select the optimum number of knots, we use the Akaike information criterion (AIC). In this simulation study, we consider the cases $p_1=p_2=3,4,5,6$ and selected the model with best value of AIC. \\

It is noted that both SSTQR (Bayes) and SSTQR (ML) perform generally better than GBSTQR. SSTQR (Bayes) performs slightly better than SSTQR (ML) which beats the GBSTQR method in performance. It should be also noted that the optimization in the ML estimation can be considerably more challenging than obtaining samples since the negative likelihood is possibly non-convex with multiple local maxima. %might under-perform based on the technique and hardness of the optimization problem. Since the negative likelihood is (possibly) non-convex (since it does not have any closed form, its not possible to verify analytically), the likelihood would have possibly multiple maximums. Hereby, depending on the starting point and the algorithm of optimization, we might end up finding different ML estimates. In terms of optimization, it is a challenging problem since the negative likelihood is not only non-convex but it can be categorized as a black-box function due to the absence of closed form. 
To find the ML estimate, we solve 250-1024 dimensional constrained black-box optimization problem (for $p_1,p_2=3,4,5,6$). 
%Under this scenario, although we have used the most convenient optimization algorithm (to the best of our knowledge), due to the hardness of the problem, it is possible to yield relatively poor estimate compared to the corresponding Bayes estimate. Despite the fact that SSTQR (ML) could not out-perform the SSTQR (Bayes) estimates, it should be noted that SSTQR (ML) is good enough to perform more or less better than GBSTQR method.

\section{Analysis of ozone concentration data of the US}
\label{ozone_us}
 Ozone concentration data of the US over the last several years can be found at the US EPA website\footnote{Source \url{https://www3.epa.gov/airquality/airdata/ad_data.html}}. The yearly averages of the daily maximum of observed hourly ozone concentration values between 9:00 AM and 8:00 PM (Daily 1-hour maximum average ozone concentration) and the daily maximum of 8 hour running average of observed hourly ozone concentration values (Daily 8-hour maximum average ozone concentration) have been collected at around 1629 sites all over the US in the last several years. 
As mentioned in the EPA website\footnote{Source\url{https://www3.epa.gov/region1/airquality/avg8hr.html}}, in 1997, EPA replaced the previous 1-hour ozone standard with 8-hour standard (which is supposed to be more protective) at a level of 84 parts per billion (ppb). 8-hour primary standard is considered to be met at any given site if the 3-year mean of the annual fourth-highest daily maximum 8-hour average ozone concentration is less than or equal to the proposed level, i.e., 84 ppb as per EPA standard announced in 1997. Later in 2008, EPA changed the 8-hour ozone concentration standard to 75 ppb and in 2015, it was decreased to 70 ppb  due to emergence of more scientific evidence regarding the effects of ozone on public health and welfare. As mentioned in a news article \cite{Barboza2015}, environmentalists and health advocacy groups including the American Lung Association endorsed 8-hour ozone standard to be 60 ppb. But \cite{Barboza2015} mentioned as per EPA data from 2014, more than 40 million people, or in other words, about 1 in every 8 people of the US lives in the counties with air pollution levels exceeding the previous ozone standard of 75 ppb.\\

Here an analysis of the spatio-temporal trend of daily 1-hour maximum average and daily 8-hour maximum average ozone concentration for the last 10 years, i.e., during the period 2006-2015, has been provided using the proposed method. It should be noted that in the given dataset, each site does not have the data for all the years considered while some of the sites have multiple observations recorded at the same year measured using different measuring instruments. With the proposed method, these cases can be handled automatically without any further required modifications. For the analysis, the Markov Chain Monte Carlo (MCMC) has been initialized at the SSTQR (ML) estimate obtained using the GCDVSMS algorithm proposed in \cite{DAS_3_2016} with the values of the tuning parameters as mentioned in the earlier section. For this analysis, 10000 iterations have been performed disregarding the first 1000 iterations as burn-in. The number of knots of the B-spline basis functions has been chosen based on the AIC criterion as mentioned in Section \ref{sim_study}.\\
%Ozone levels have declined by about one-third nationwide since 1980 as a result of regulations targeting emissions from cars, factories, consumer products and other sources of pollutants, according to the EPA \cite{Barboza2015}.

\begin{figure}
    \centering
    \includegraphics[width=.6\linewidth]{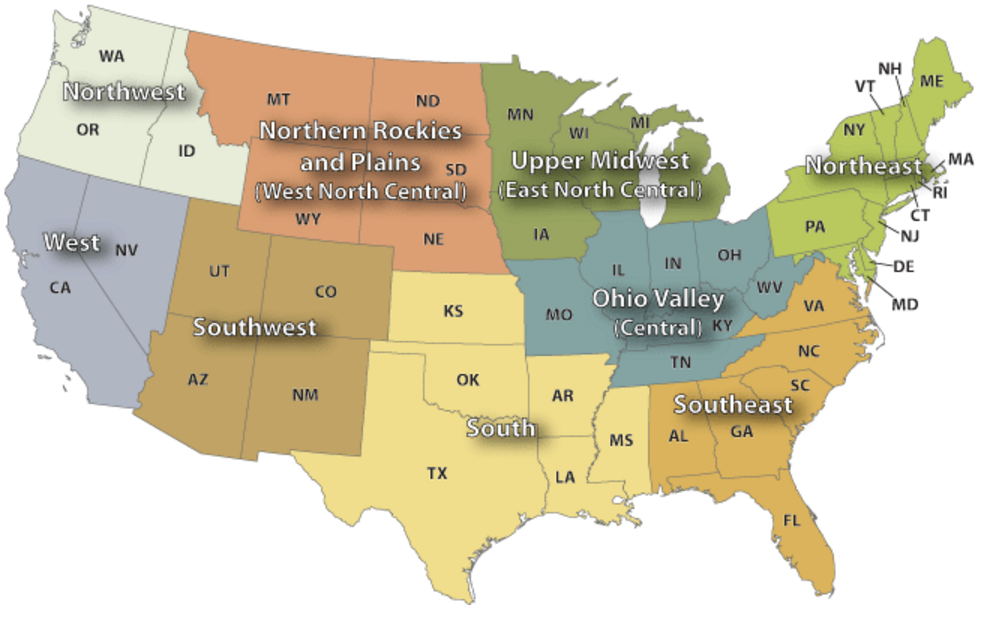} 
    \label{fig:us_climate_regions} 
  \caption{Climate region-wise division of the US.}
  \label{figfigus} 
\end{figure}

\begin{figure} 
  \begin{subfigure}[b]{.75\linewidth}
    \centering
    \includegraphics[width=0.99\linewidth]{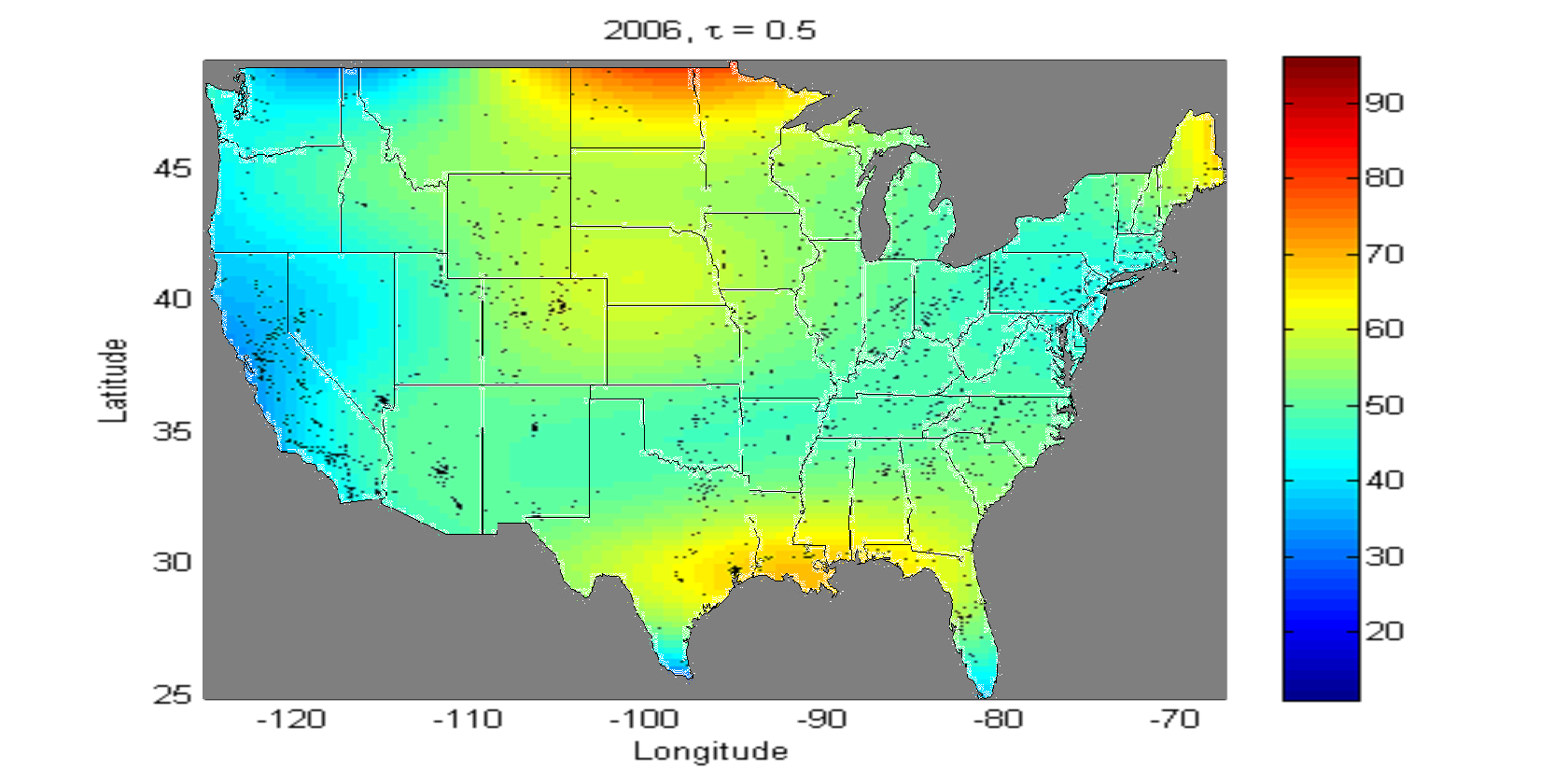} 
    \caption{Year : $2006, \tau = 0.5$} 
    \label{fig:spl_1_50_06} 
  \end{subfigure}\\
  \begin{subfigure}[b]{.75\linewidth}
    \centering
    \includegraphics[width=0.99\linewidth]{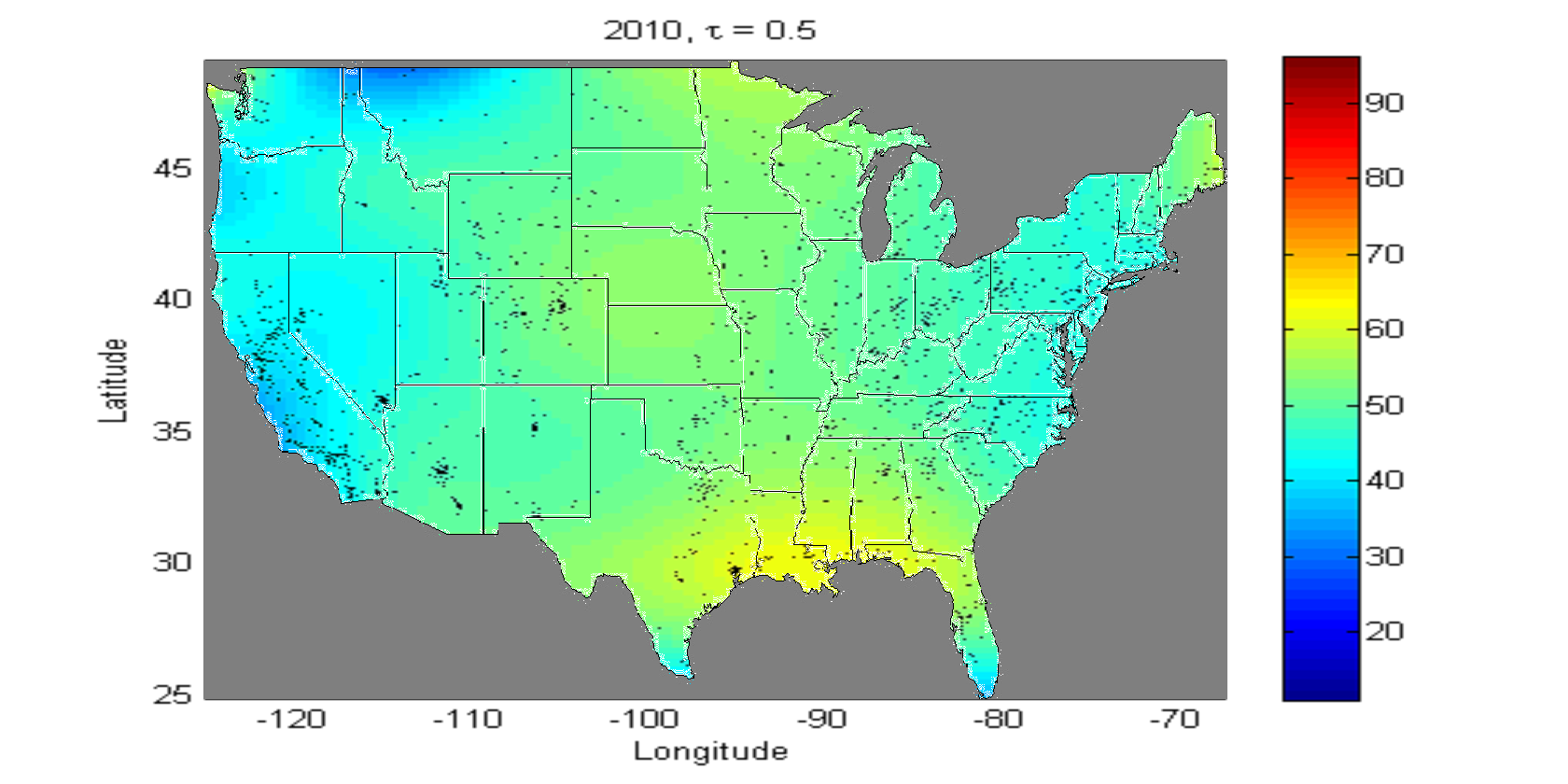} 
    \caption{Year : $2010, \tau = 0.5$} 
    \label{fig:spl_1_50_10} 
  \end{subfigure}\\
   \begin{subfigure}[b]{.75\linewidth}
    \centering
    \includegraphics[width=0.99\linewidth]{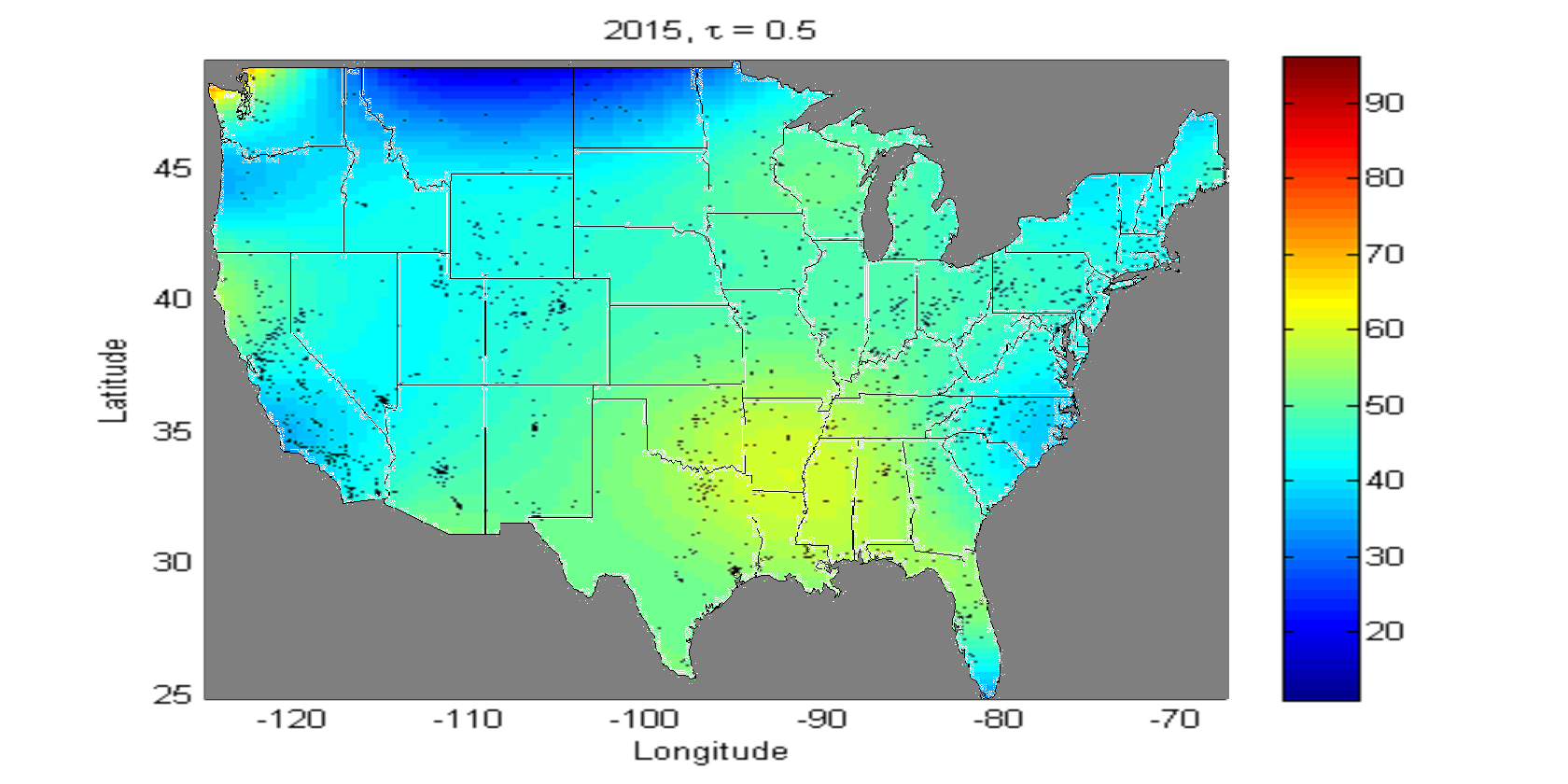} 
    \caption{Year : $2015, \tau = 0.5$}  
    \label{fig:spl_1_50_15} 
  \end{subfigure} \\
  \caption{Daily 1-hour maximum average ozone concentration (in ppb) of the US in 2006, 2010 and 2015 at $\tau = 0.5$. The dots denote weather stations where data have been collected.}
  \label{figfig1} 
\end{figure}

\begin{figure}
  \begin{subfigure}[b]{.75\linewidth}
    \centering
    \includegraphics[width=0.99\linewidth]{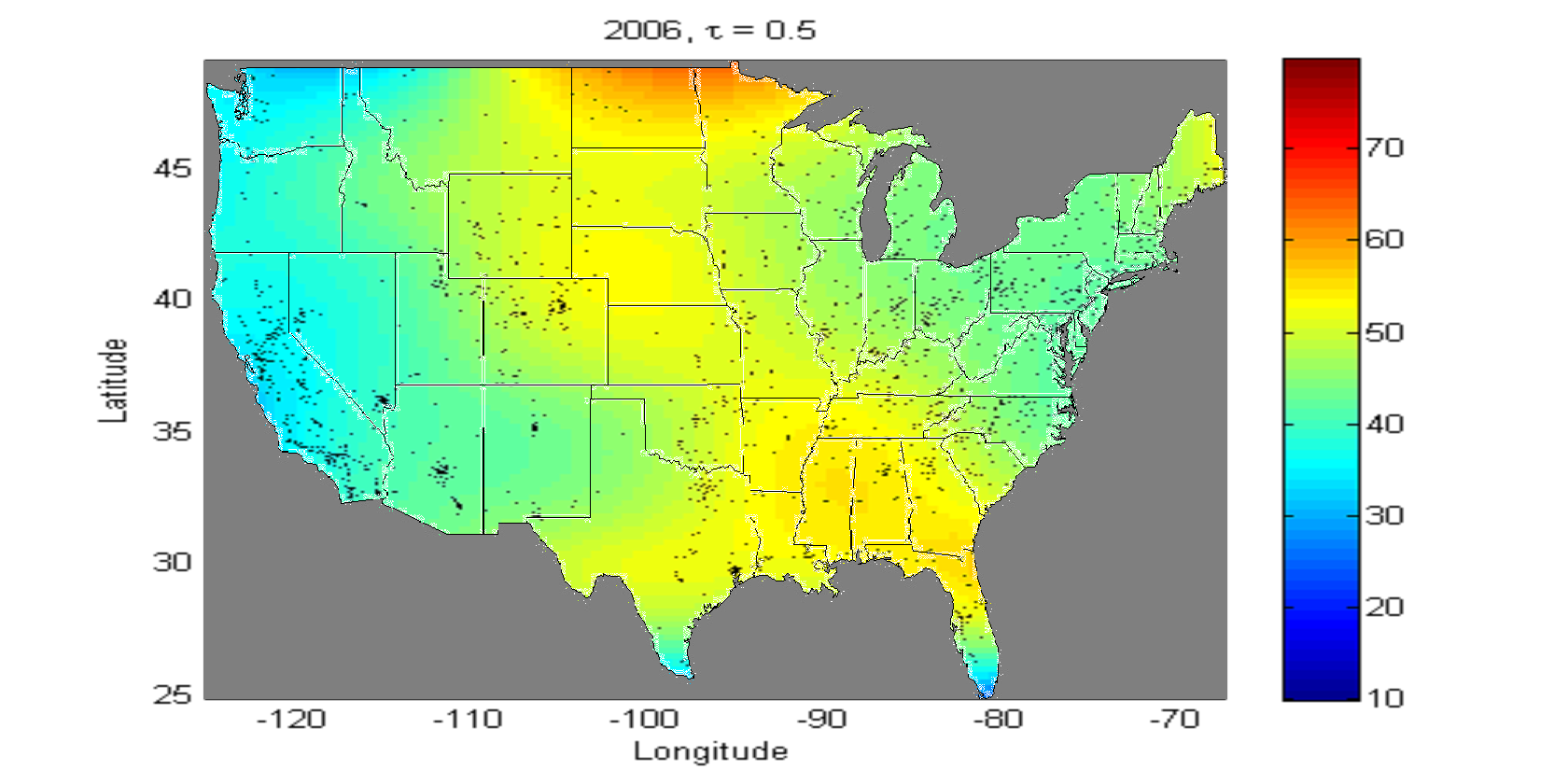} 
    \caption{Year : $2006, \tau = 0.5$} 
    \label{fig:spl_2_50_06} 
  \end{subfigure}\\
  \begin{subfigure}[b]{.75\linewidth}
    \centering
    \includegraphics[width=0.99\linewidth]{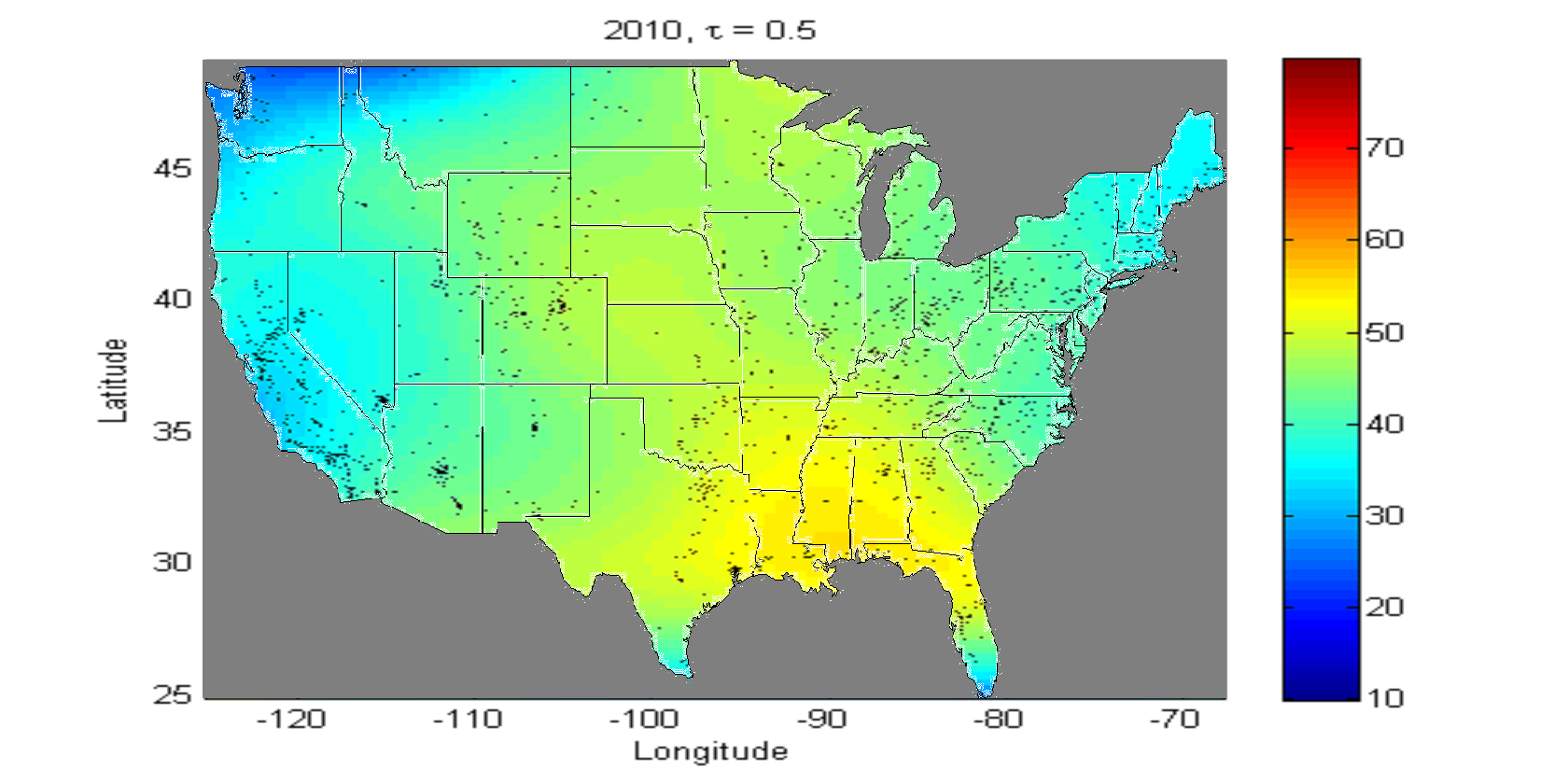} 
    \caption{Year : $2010, \tau = 0.5$} 
    \label{fig:spl_2_50_10} 
  \end{subfigure}\\
   \begin{subfigure}[b]{.75\linewidth}
    \centering
    \includegraphics[width=0.99\linewidth]{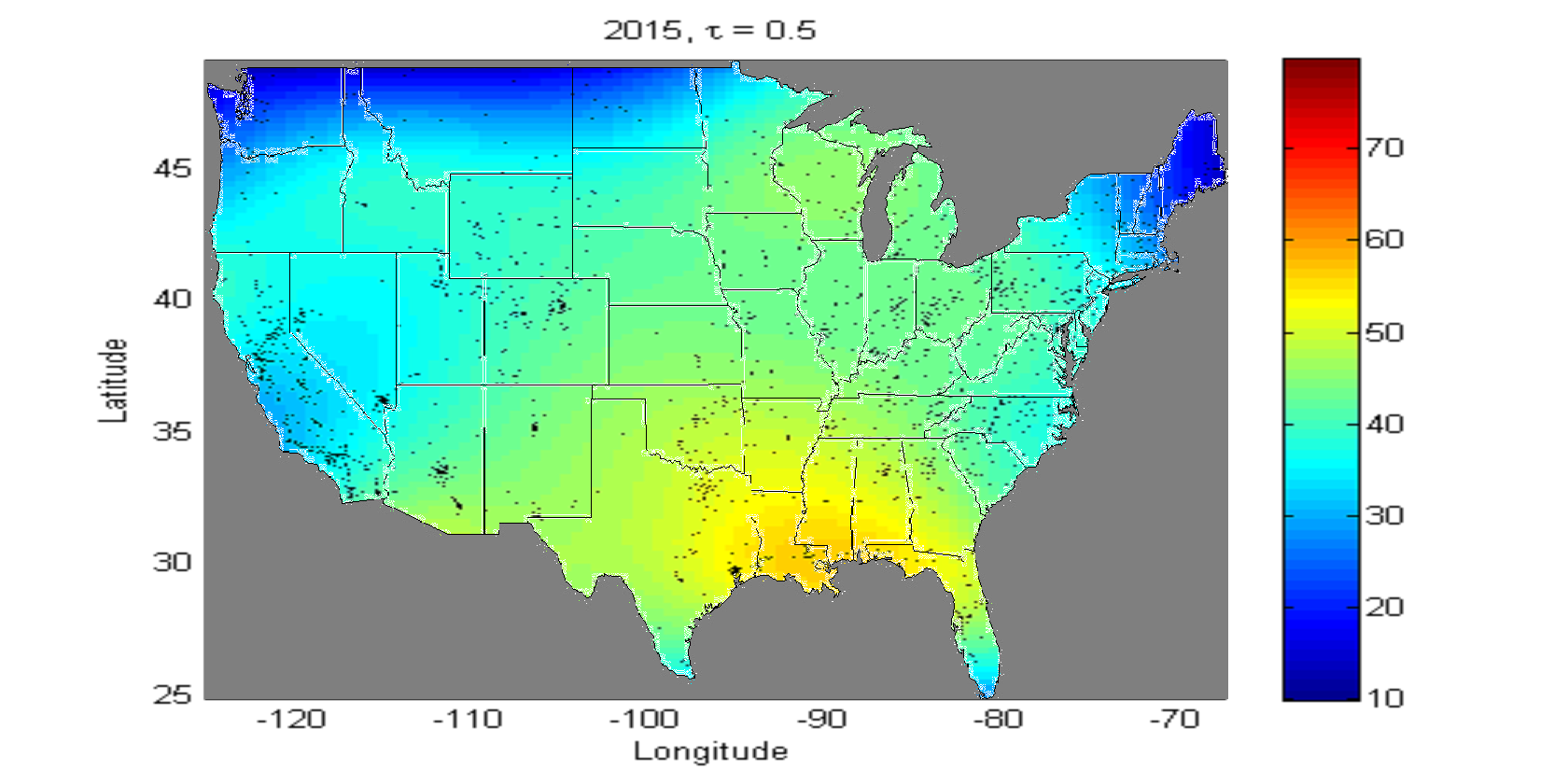} 
    \caption{Year : $2015, \tau = 0.5$}  
    \label{fig:spl_2_50_15} 
  \end{subfigure} \\
  \caption{Daily 8-hour maximum average ozone concentration (in ppb) of the US in 2006, 2010 and 2015 at $\tau = 0.5$. The dots denote weather stations where data have been collected.}
  \label{figfig3} 
\end{figure}

\begin{figure} 
  \begin{subfigure}[b]{.75\linewidth}
    \centering
    \includegraphics[width=0.99\linewidth]{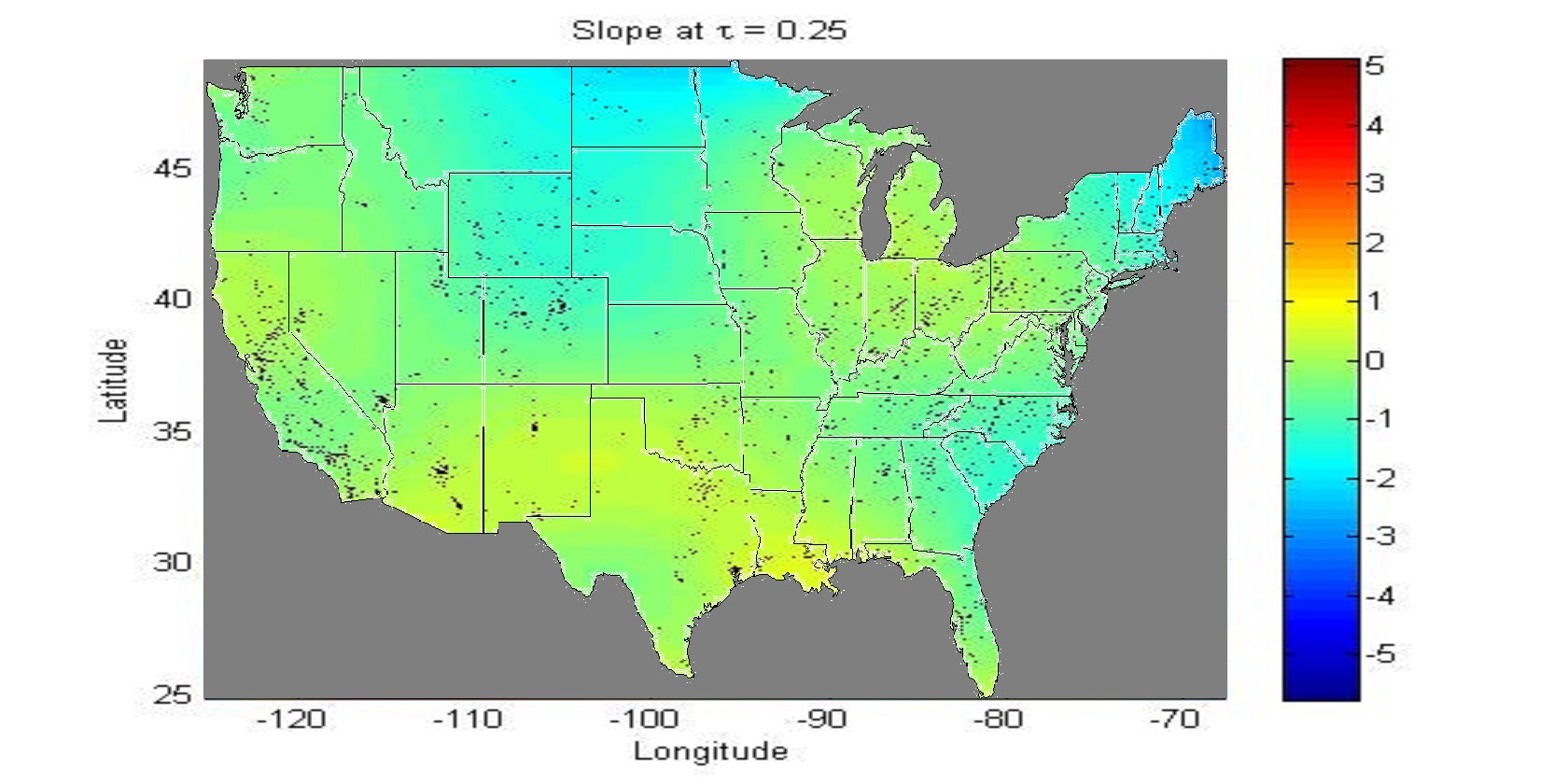} 
    \caption{Slope at $\tau = 0.25$} 
    \label{fig:SLOPE_25_8hr_BETTER} 
  \end{subfigure}\\
  \begin{subfigure}[b]{.75\linewidth}
    \centering
    \includegraphics[width=0.99\linewidth]{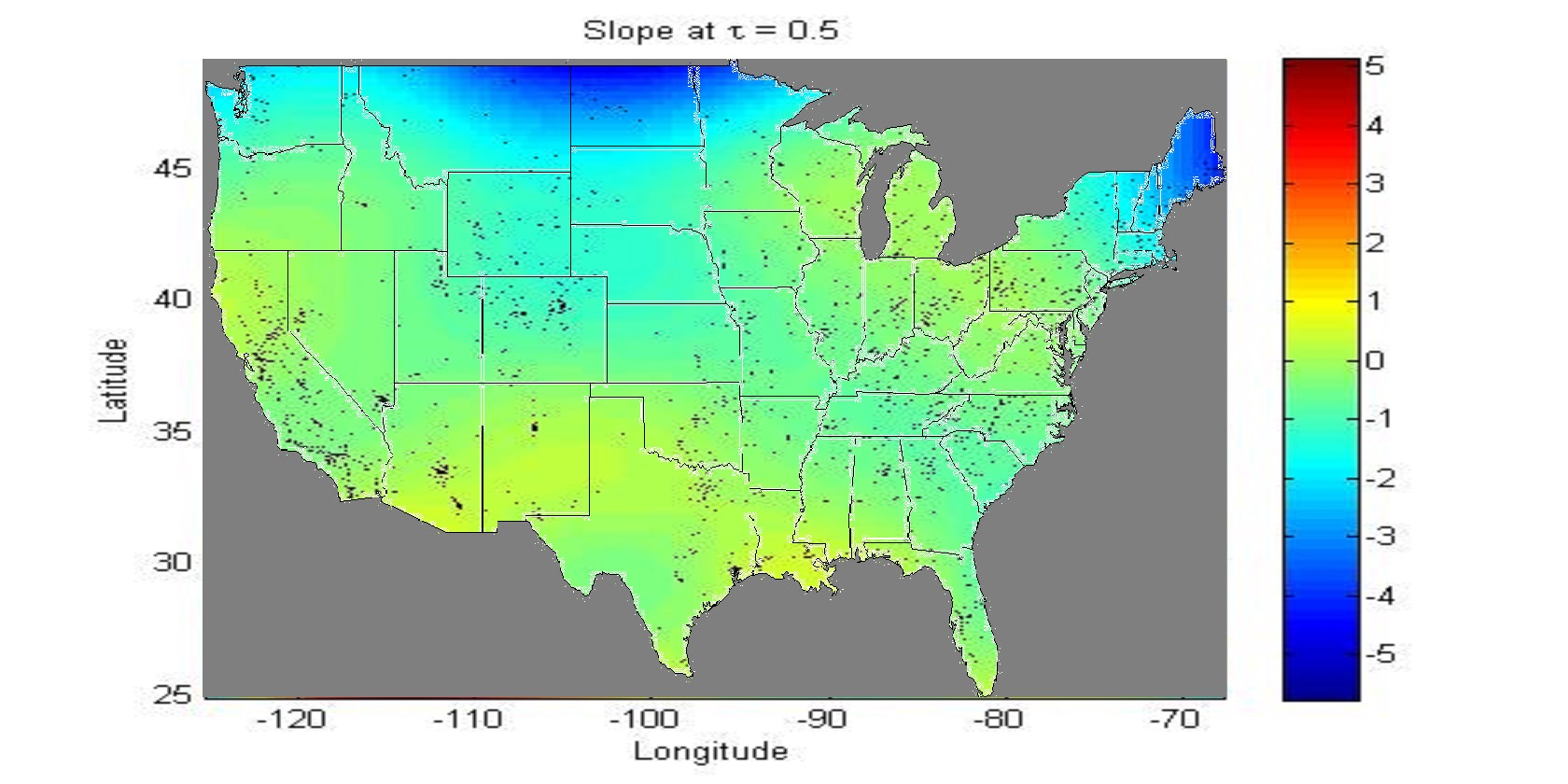} 
    \caption{Slope at $\tau = 0.5$} 
    \label{fig:SLOPE_50_8hr_BETTER} 
  \end{subfigure}\\
   \begin{subfigure}[b]{.75\linewidth}
    \centering
    \includegraphics[width=0.99\linewidth]{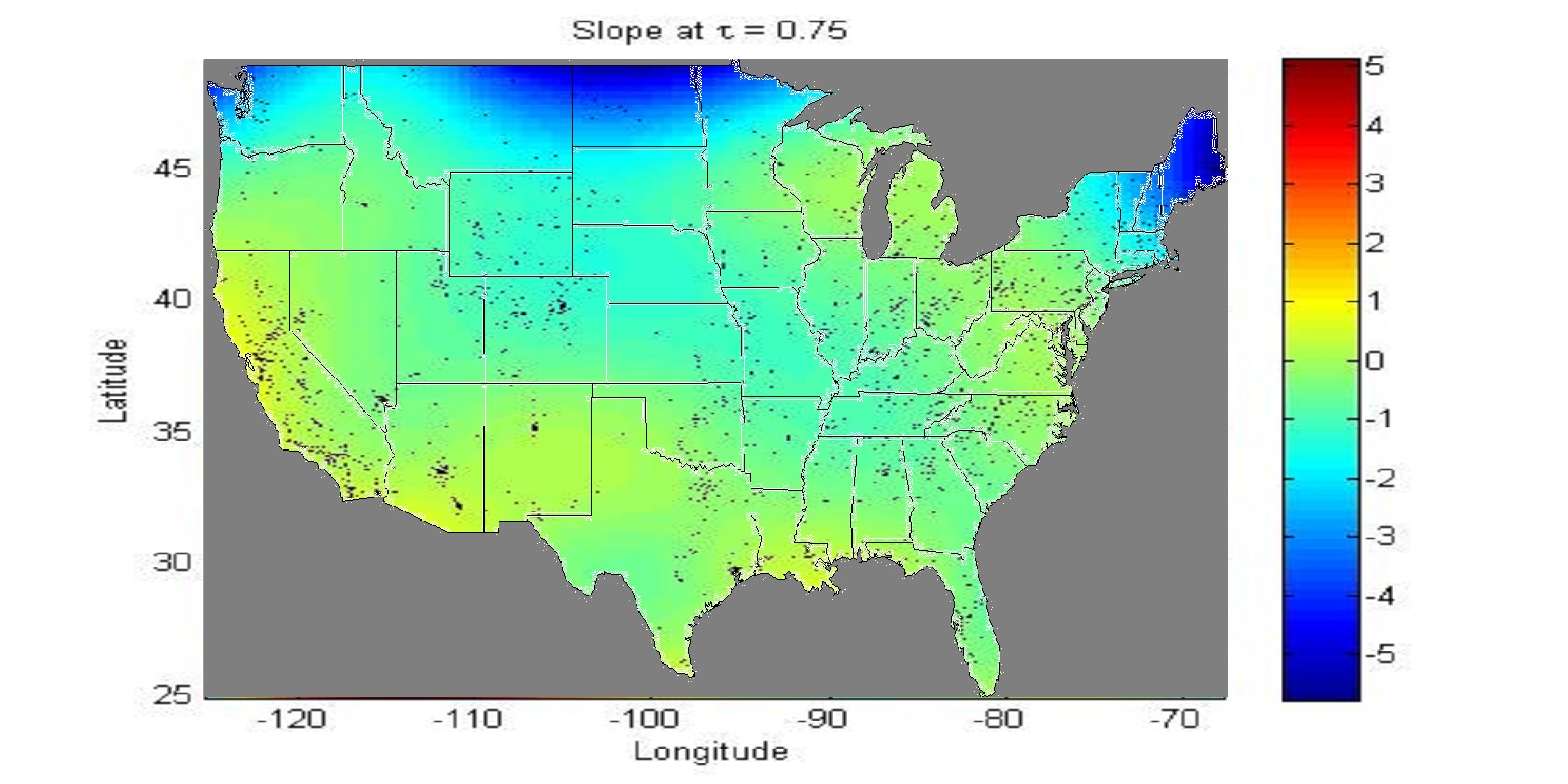} 
    \caption{Slope at $\tau = 0.75$}  
    \label{fig:SLOPE_75_8hr_BETTER} 
  \end{subfigure} \\
  \caption{Yearly rate of change of daily 8-hour maximum average ozone concentration (in ppb/year) of the US at $\tau = 0.25, 0.5, 0.75$. The dots denote weather stations where data have been collected.}
  \label{figfig4} 
\end{figure}

\begin{figure} 
  \begin{subfigure}[b]{.75\linewidth}
    \centering
    \includegraphics[width=0.99\linewidth]{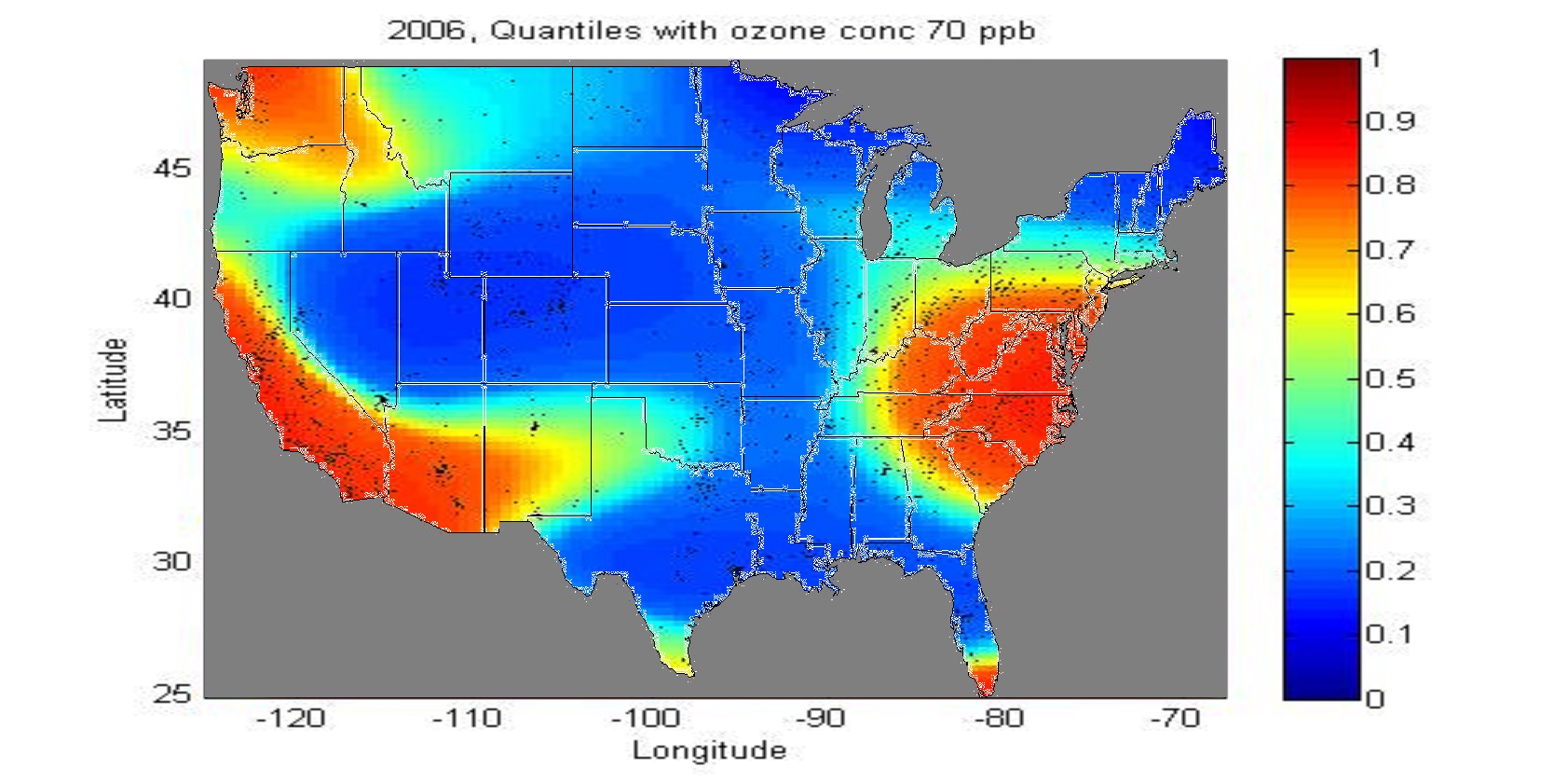} 
    \caption{2006, quantiles at 4th highest 8-hr max ozone conc. 70 ppb} 
    \label{fig:quantiles_8hr_1_better} 
  \end{subfigure}\\
  \begin{subfigure}[b]{.75\linewidth}
    \centering
    \includegraphics[width=0.99\linewidth]{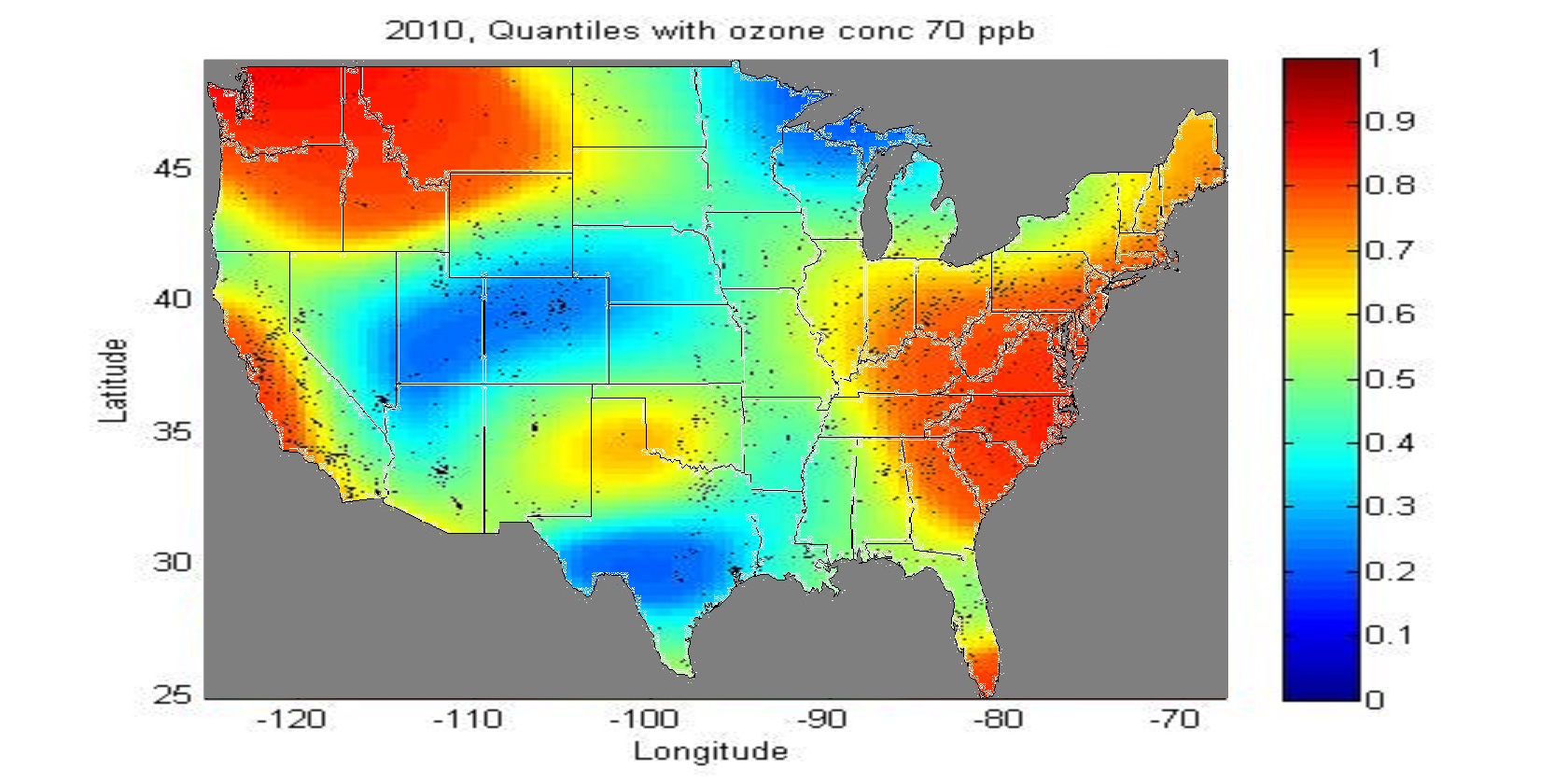} 
    \caption{2010, quantiles at 4th highest 8-hr max ozone conc. 70 ppb} 
    \label{fig:quantiles_8hr_2_better} 
  \end{subfigure}\\
   \begin{subfigure}[b]{.75\linewidth}
    \centering
    \includegraphics[width=0.99\linewidth]{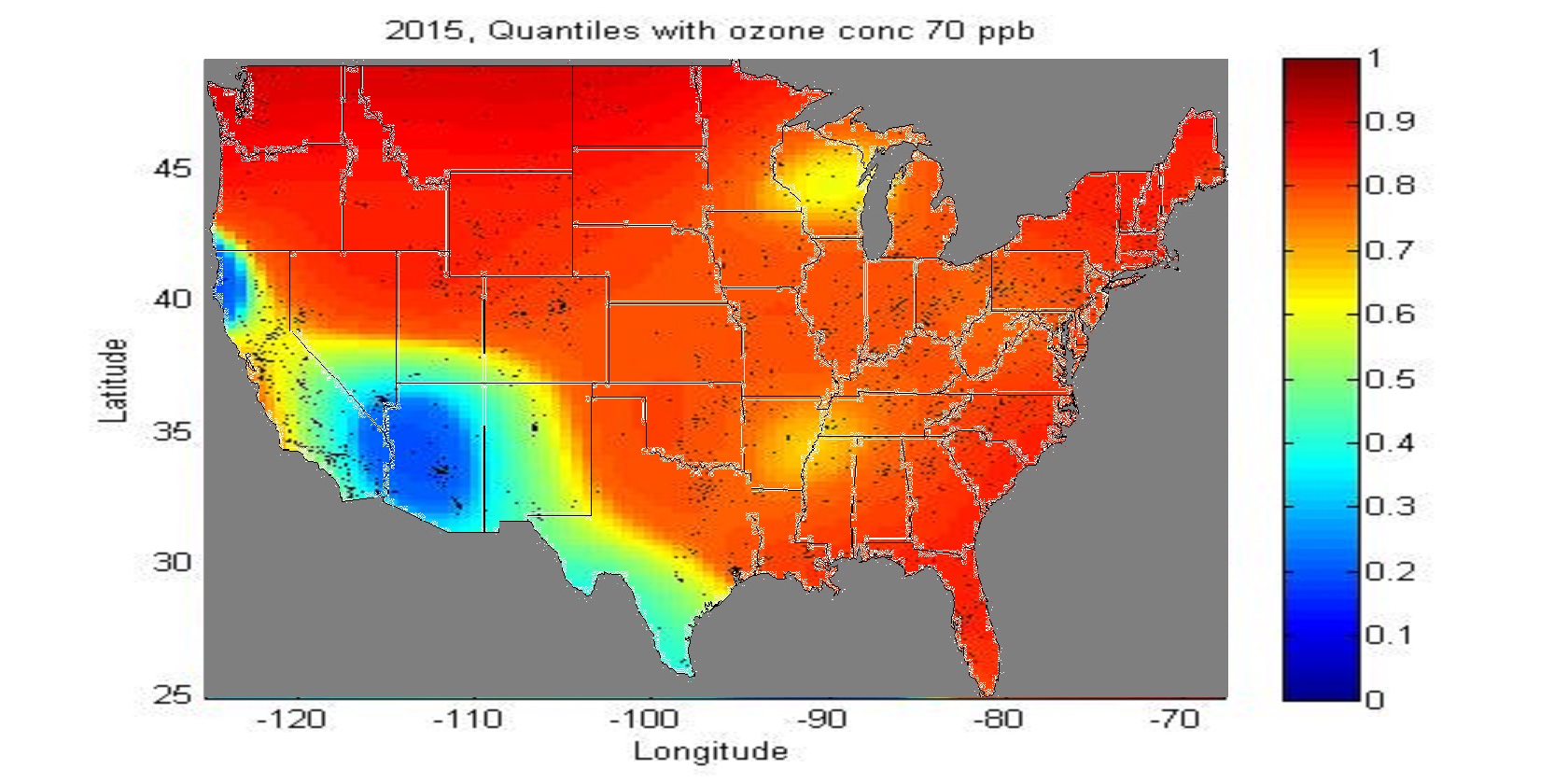} 
    \caption{2015, quantiles at 4th highest 8-hr max ozone conc. 70 ppb}  
    \label{fig:quantiles_8hr_3_better} 
  \end{subfigure} \\
  \caption{Quantiles at which 4th highest daily 8-hour maximum ozone concentration (in ppb) of the US is 70 ppb  in 2006, 2010 and 2015.}
  \label{figfig4a} 
\end{figure}

For the years 2006, 2010 and 2015 the daily 1-hour maximum and 8-hour maximum average ozone concentration are plotted over the US at the median level. It is noted that in the most of the parts of the US the daily 1-hour maximum average ozone concentration has decreased over time. Specifically the decreasing trend of 1-hour concentration is noticeable in the West North Central, Upper Midwest, Northeast and southern part of the South climate regions (see Figure \ref{figfigus}, \ref{figfig1}). The daily 8-hour maximum average ozone concentration level also has an overall decreasing trend over the time period considered. The 8-hour concentration has decreased quite a bit in the upper Northwest and West North Central regions. Besides, it has also decreased in the Northeast, South and Southeast regions (see Figures \ref{figfig3}, \ref{figfig4}). In Figure \ref{figfig4a} we also show the quantile levels across the US where the 4th highest daily 8-hour maximum ozone concentration is 70 ppb. It is noted that over the last ten years the 4th highest daily 8-hour maximum ozone concentration in the US has generally a decreasing trend since 70 ppb ozone concentration is attained at relatively higher quantiles in the latter years.  \\

\begin{figure} 
  \begin{subfigure}[b]{.75\linewidth}
    \centering
    \includegraphics[width=0.99\linewidth]{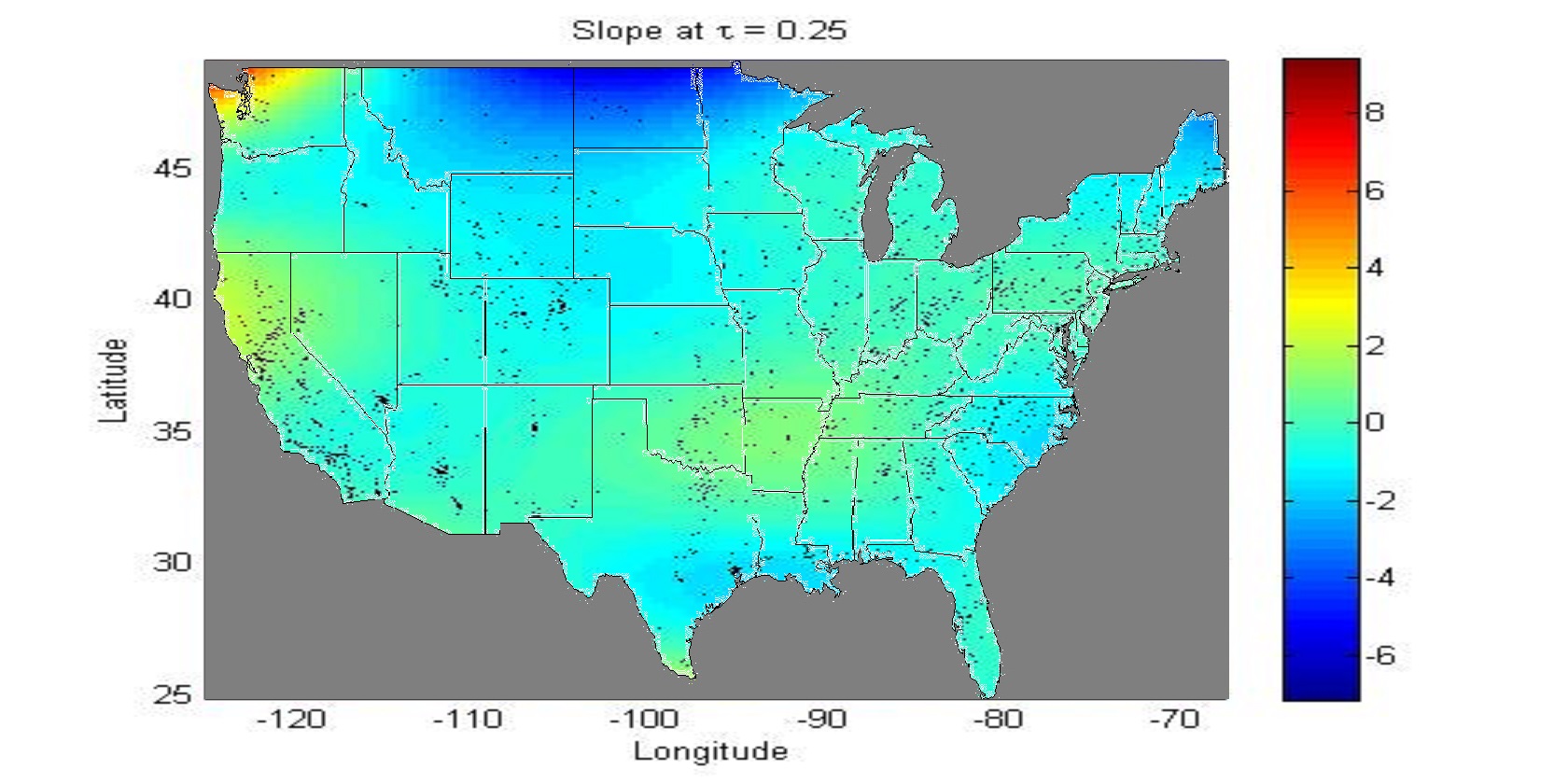} 
    \caption{Slope at $\tau = 0.25$} 
    \label{fig:SLOPE_25_1hr_BETTER} 
  \end{subfigure}\\
  \begin{subfigure}[b]{.75\linewidth}
    \centering
    \includegraphics[width=0.99\linewidth]{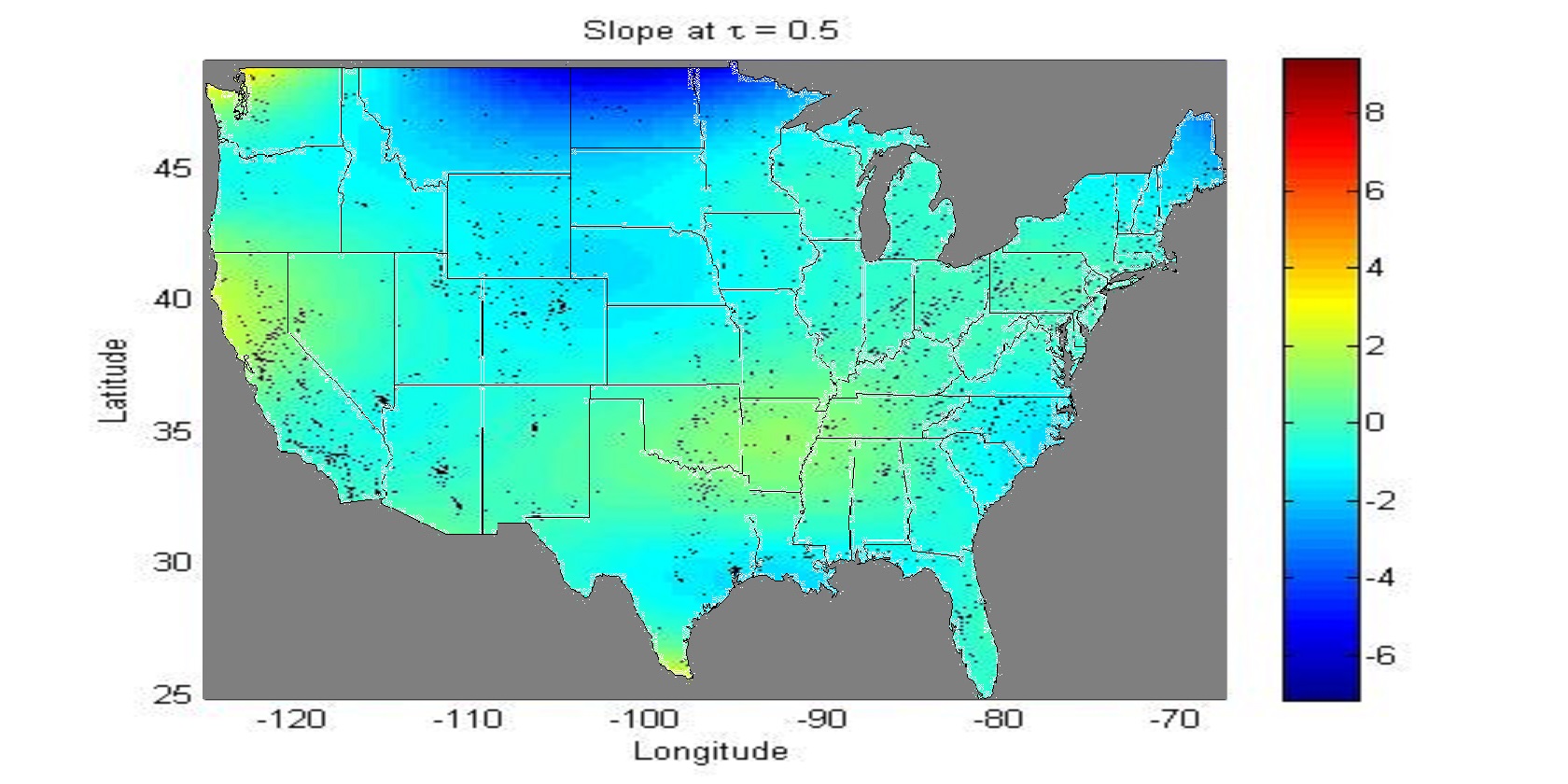} 
    \caption{Slope at $\tau = 0.5$} 
    \label{fig:SLOPE_50_1hr_BETTER} 
  \end{subfigure}\\
   \begin{subfigure}[b]{.75\linewidth}
    \centering
    \includegraphics[width=0.99\linewidth]{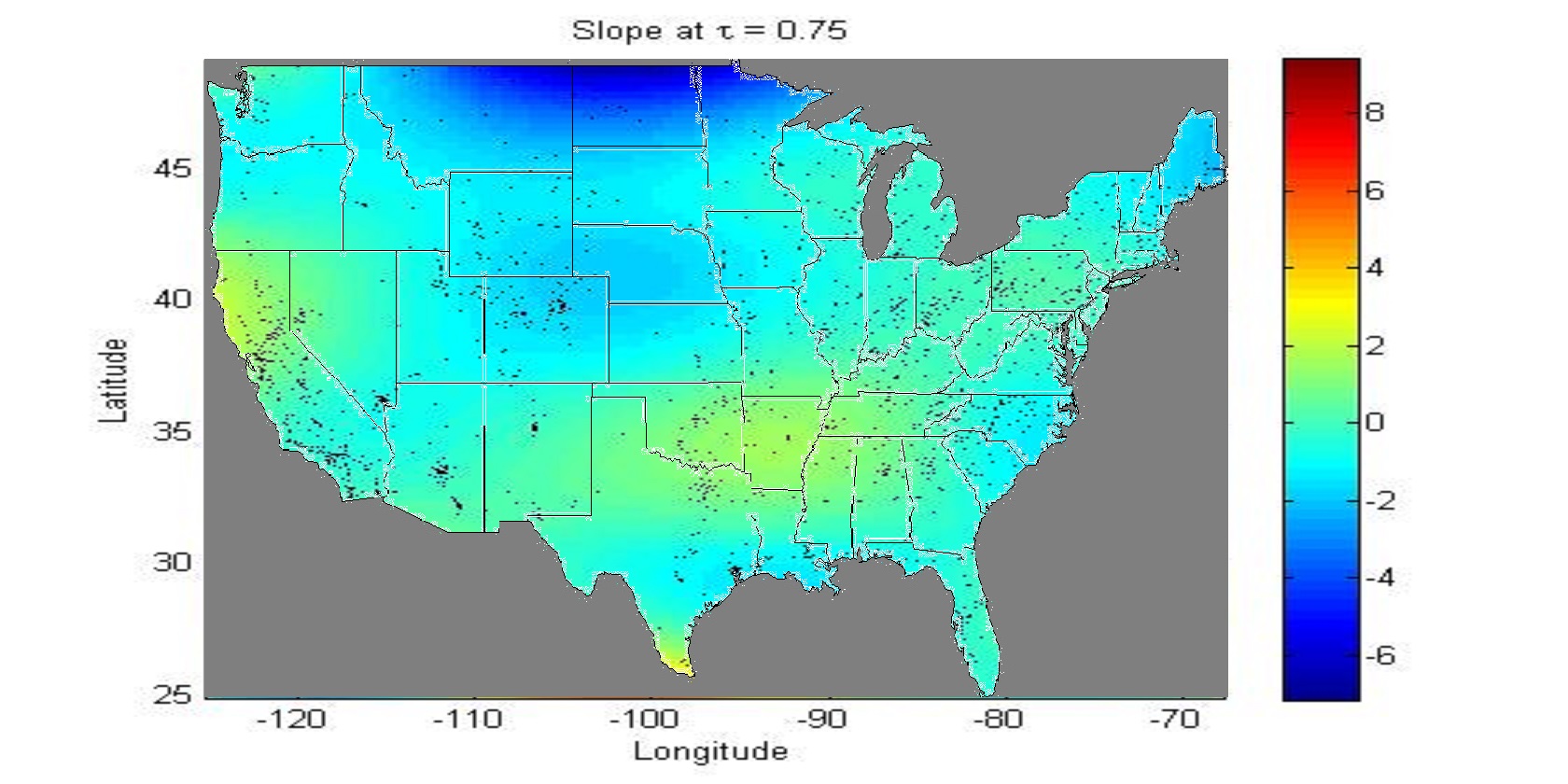} 
    \caption{Slope at $\tau = 0.75$}  
    \label{fig:SLOPE_75_1hr_BETTER} 
  \end{subfigure} \\
  \caption{Yearly rate of change of daily 1-hour maximum average ozone concentration (in ppb/year) of the US at $\tau = 0.25, 0.5, 0.75$. The dots denote weather stations where data have been collected.}
  \label{figfig2} 
\end{figure}

\section{Analysis of ozone concentration data of California}
\label{ozone_cali}
As mentioned in \cite{Barboza2015}, California has the worst smog in the US and it does not meet the existing smog limits. It also says that the air quality is the worst in the inland valleys, where pollution from vehicles and factories yields ozone in the presence of sunlight and that ozone is blown and trapped against the mountains. Some of the areas in California are so polluted that as per EPA, they are expected to meet the standards only by 2037, needing 12 more years to recover unlike the rest of the nation. According to the state Air Resources Board, about one-third population of California live in communities where the pollution level exceeds federal standards. As mentioned in the California Environmental Protection Agency site\footnote{Source\url{https://www.arb.ca.gov/research/aaqs/caaqs/ozone/ozone.htm}}, in April 2005, the Air Resources Board retained the previous 1-hour ozone standard of 90 ppb and set a new 8-hour standard of 70 ppb. This ozone standard review was also mandated by the Children's Environmental Health Protection Act.\\

\begin{figure}
    \centering
    \includegraphics[width=0.4\linewidth]{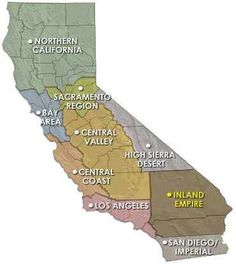}  
  \caption{Region-wise division of California.}
  \label{figfigcali} 
\end{figure}

\begin{figure}
  \begin{subfigure}[b]{0.5\linewidth}
    \centering
    \includegraphics[width=0.99\linewidth]{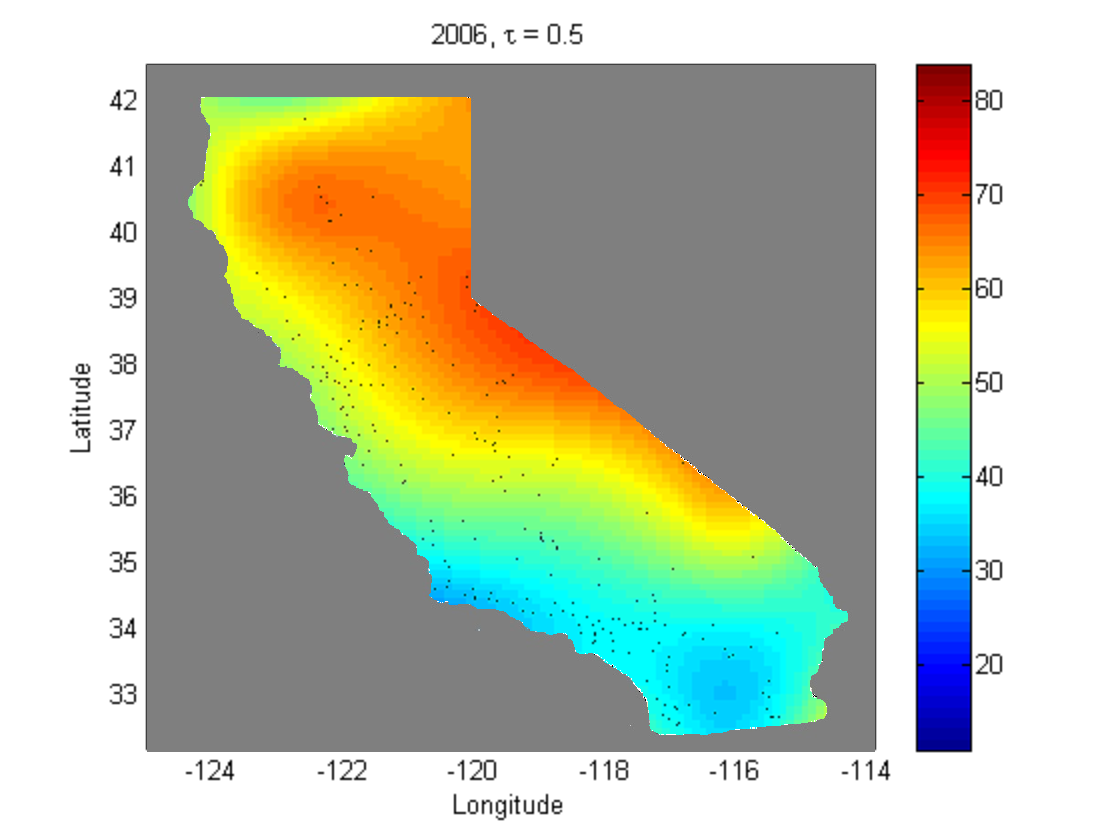} 
    \caption{Year : $2006, \tau = 0.5$} 
    \label{fig:Cali_1_50_06} 
  \end{subfigure}
   \begin{subfigure}[b]{0.5\linewidth}
      \centering
      \includegraphics[width=0.99\linewidth]{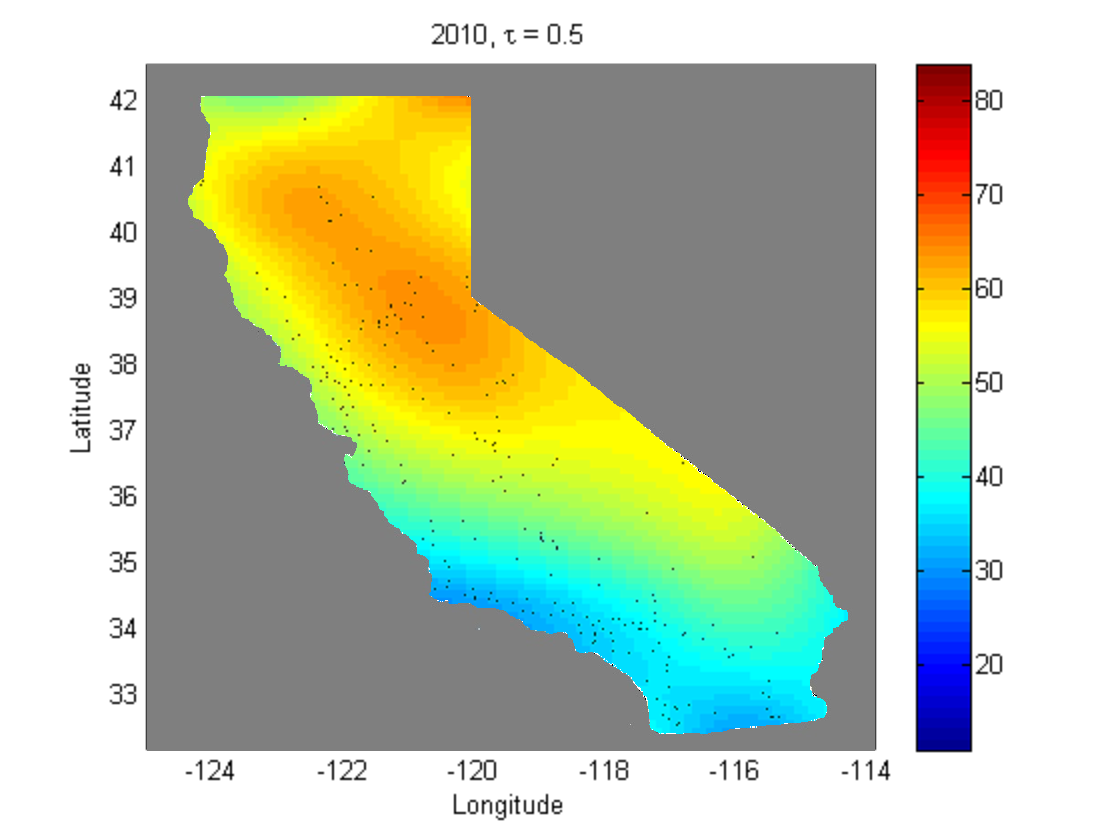} 
      \caption{Year : $2010, \tau = 0.5$} 
      \label{fig:Cali_1_50_10} 
   \end{subfigure}\\
    \begin{subfigure}[b]{0.5\linewidth}
     \centering
     \includegraphics[width=0.99\linewidth]{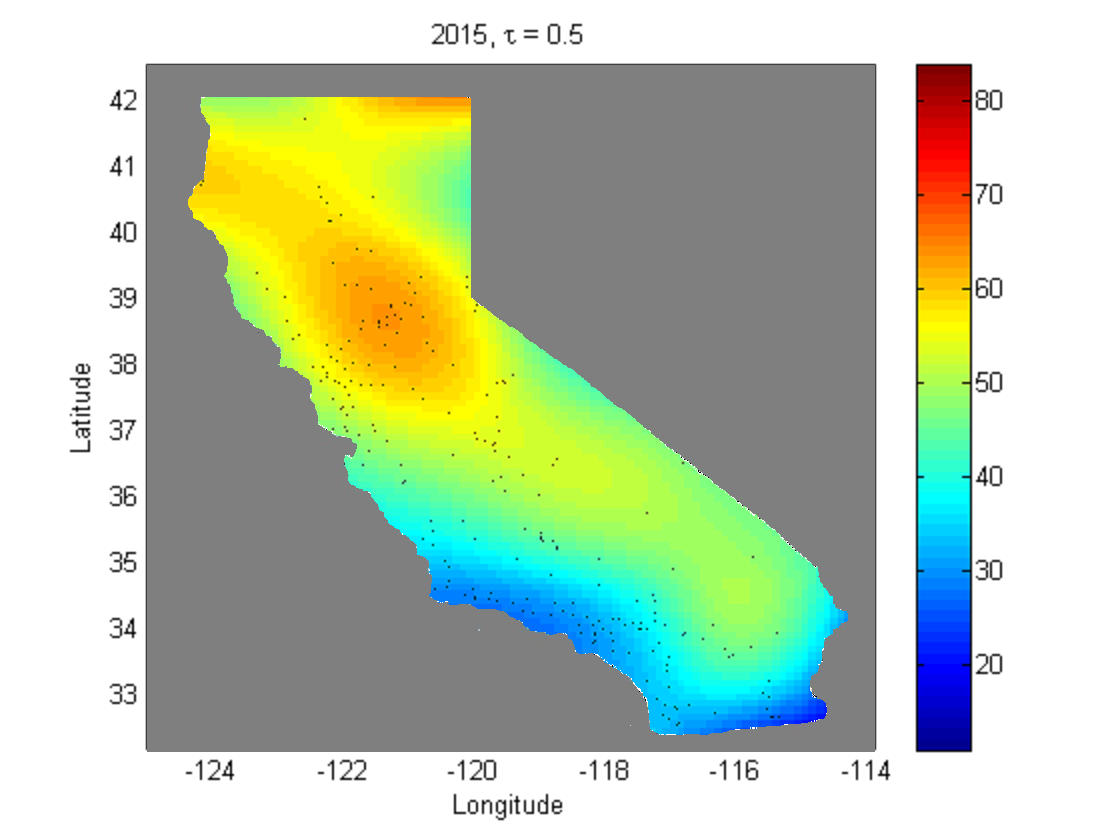} 
     \caption{Year : $2015, \tau = 0.5$}  
     \label{fig:Cali_1_50_15} 
   \end{subfigure}  
    \begin{subfigure}[b]{0.5\linewidth}
    \centering
    \includegraphics[width=0.99\linewidth]{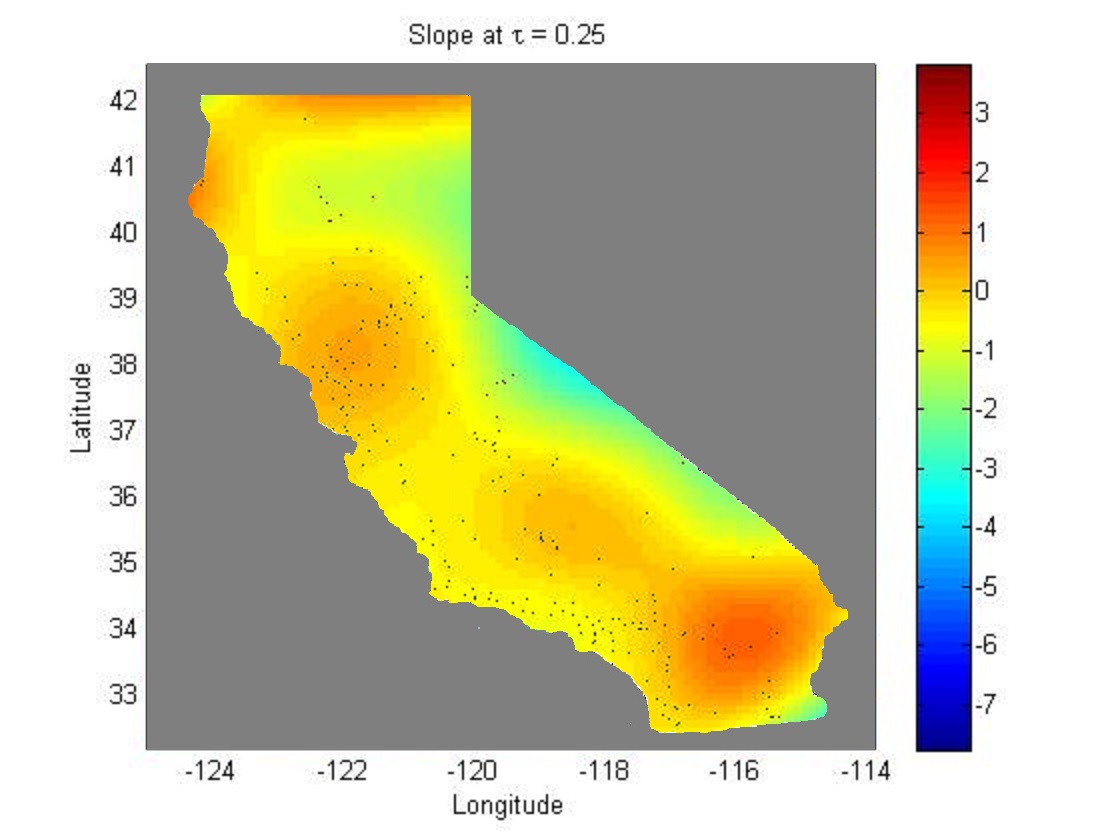} 
    \caption{Slope at $\tau = 0.25$}  
    \label{fig:CALI_SLOPE_25_1hr_BETTER} 
  \end{subfigure} \\
  \begin{subfigure}[b]{0.5\linewidth}
    \centering
    \includegraphics[width=0.99\linewidth]{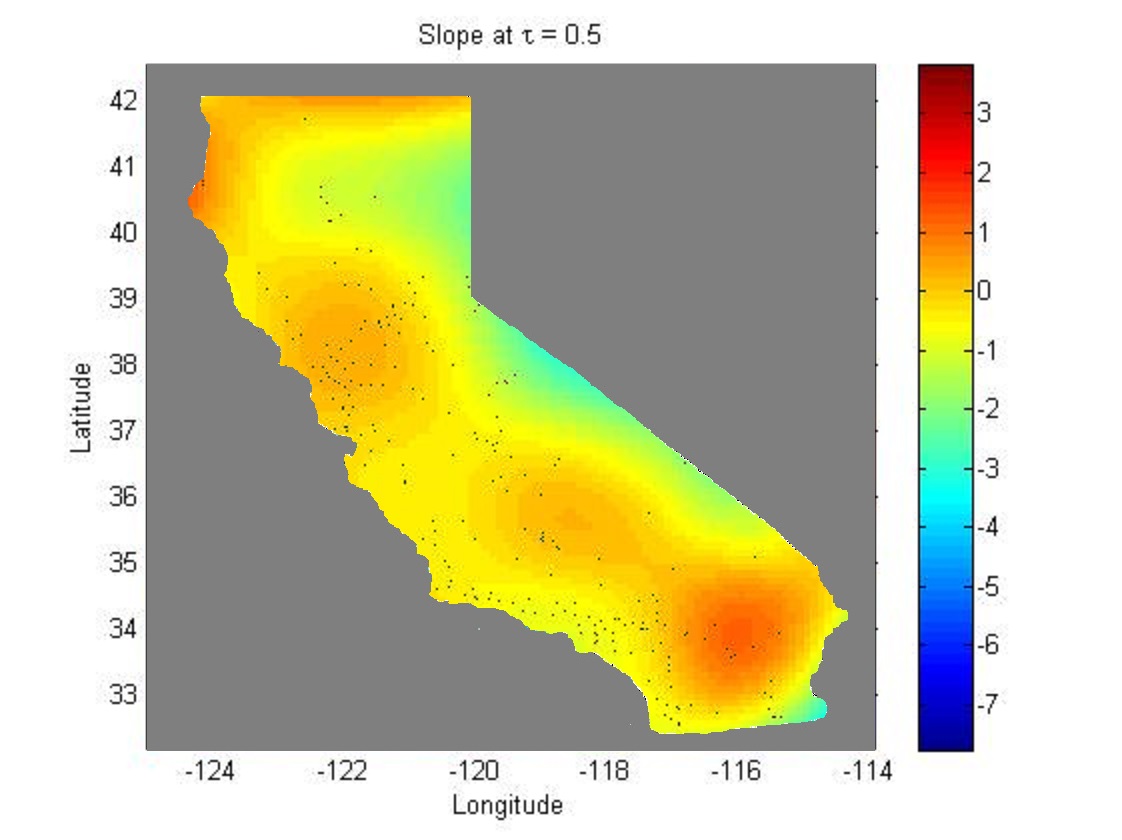} 
    \caption{Slope at $\tau = 0.5$} 
    \label{fig:CALI_SLOPE_50_1hr_BETTER} 
  \end{subfigure}
   \begin{subfigure}[b]{0.5\linewidth}
    \centering
    \includegraphics[width=0.99\linewidth]{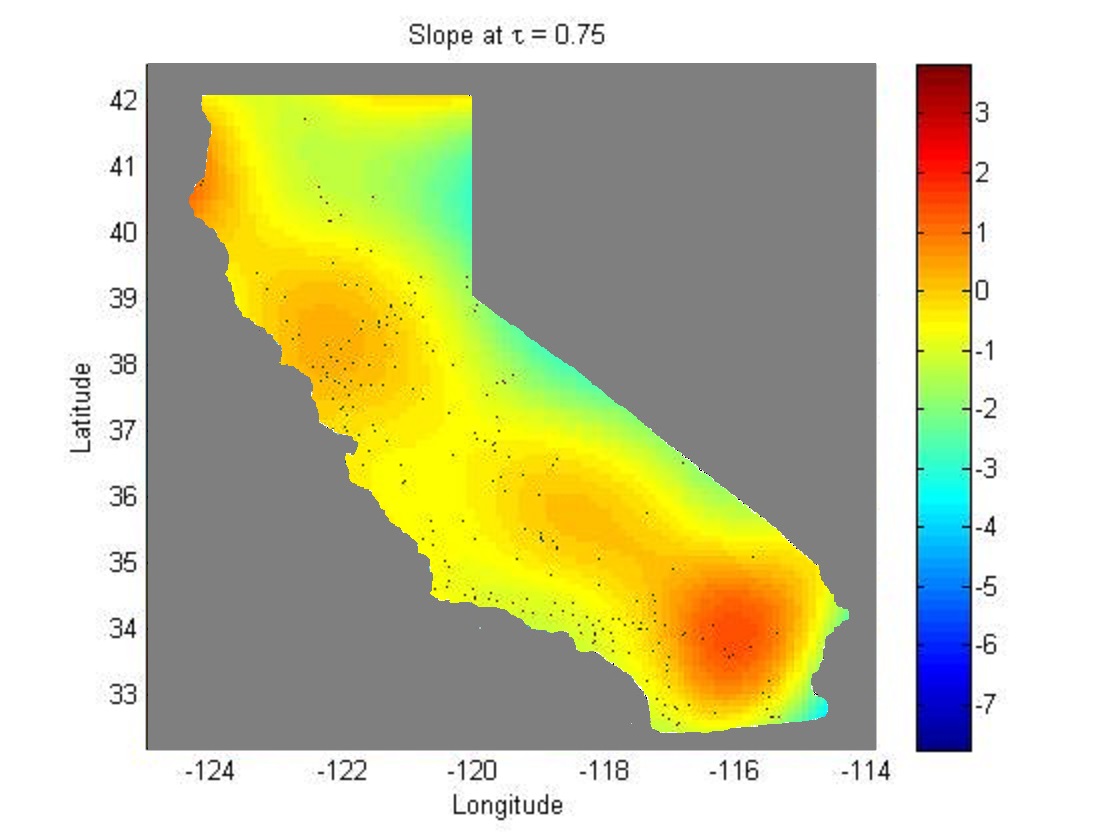} 
   \caption{Slope at $\tau = 0.75$}  
    \label{fig:CALI_SLOPE_75_1hr_BETTER} 
  \end{subfigure}
  \caption{(a-c) Daily 1-hour maximum average ozone concentration (in ppb) of California in 2006, 2010 and 2015 at $\tau = 0.5$. (d-f) Yearly rate of change of daily 1-hour maximum average ozone concentration (in ppb/year) of California at $\tau = 0.25, 0.5, 0.75$.The dots denote weather stations where data have been collected.}
  \label{figfig5} 
\end{figure}

In this section, the spatio-temporal trend of daily 1-hour maximum and 8-hour maximum average ozone concentration data of California are analyzed over the time period 2006-2015 and spatial plots are shown for the years 2006, 2010 and 2015 at $50$-th quantile level. For those three years, we also show the quantile levels across California where the 4th highest daily 8-hour maximum ozone concentration is 70 ppb and 4th highest daily 1-hour maximum ozone concentration is 90 ppb. Similar to the previous section, the SSTQR (Bayes) method is used for the analysis and SSTQR (ML) estimate is used as the starting point of the MCMC chain. In this case also 10000 iterations are performed and the first 1000 iterations are disregarded as burn-in. It is noted that overall there is a decreasing trend of daily 1-hour maximum average concentration level at both the quantile levels considered. Specifically in the Northern California, Sacramento Region, Central Valley, High Sierra Desert, Los Angeles and San Diego (see Figures \ref{figfigcali}, \ref{figfig5})  the 1-hour maximum average ozone concentration has noticeably decreased over time. It is also noted that in the Inland Empire region, it has a slightly increasing trend over time. As mentioned earlier, unlike in other parts of the US, Air Resources Board retained 1-hour ozone standard in California and its 8-hour standard has been 70 ppb for the last ten years while till September 2015, 8-hour standard for the rest of the country has been greater than or equal to 75 ppm. Thus to comply with the stricter rules, tighter controls on factories, vehicles, power plants and other emitters of smog-forming pollutants, the overall daily 1-hour maximum average ozone concentration of California shows a decreasing trend. \\

\begin{figure} 
  \begin{subfigure}[b]{0.5\linewidth}
    \centering
    \includegraphics[width=0.99\linewidth]{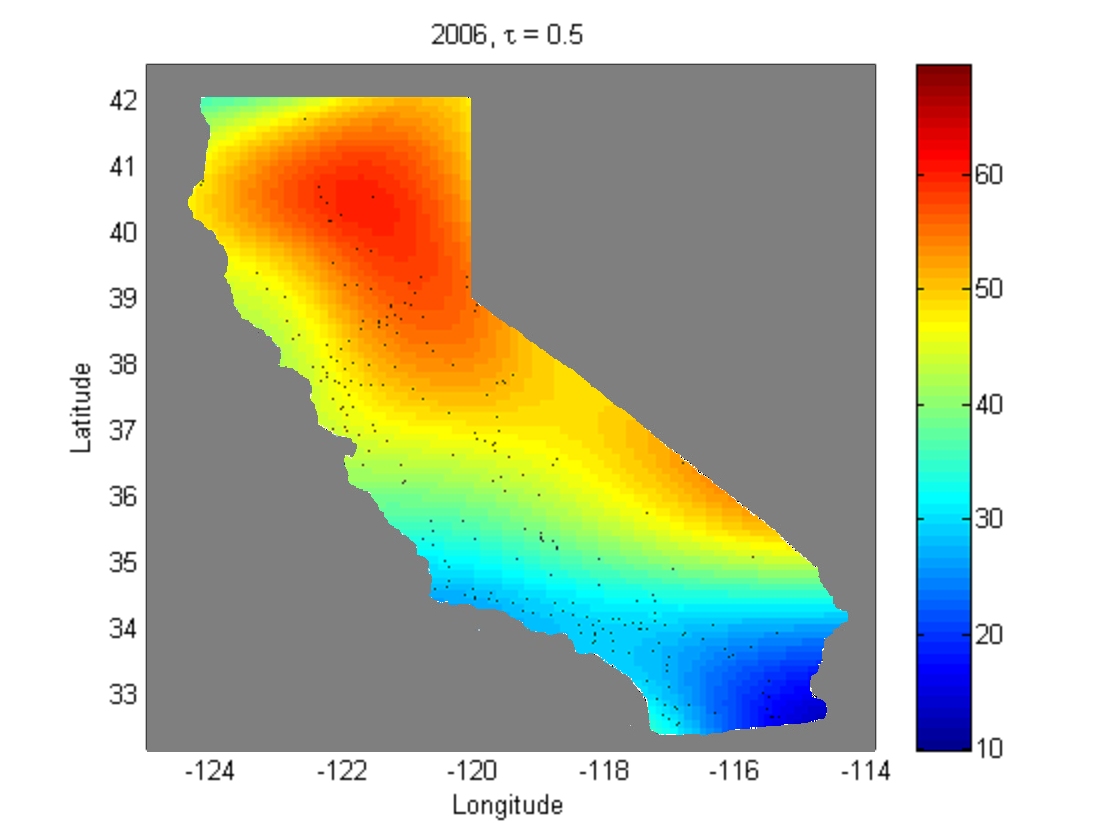} 
    \caption{Year : $2006, \tau = 0.5$} 
    \label{fig:Cali_2_50_06} 
  \end{subfigure}
  \begin{subfigure}[b]{0.5\linewidth}
    \centering
    \includegraphics[width=0.99\linewidth]{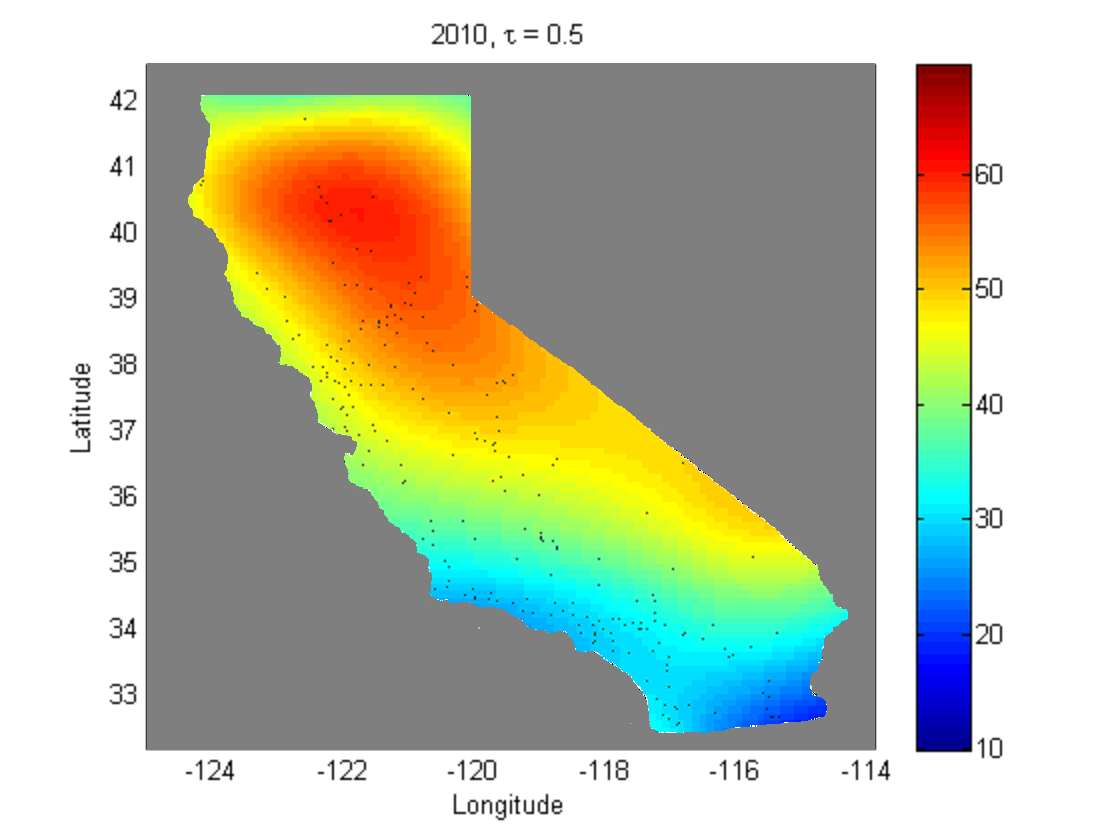} 
    \caption{Year : $2010, \tau = 0.5$} 
    \label{fig:Cali_2_50_10} 
  \end{subfigure}\\
   \begin{subfigure}[b]{0.5\linewidth}
    \centering
    \includegraphics[width=0.99\linewidth]{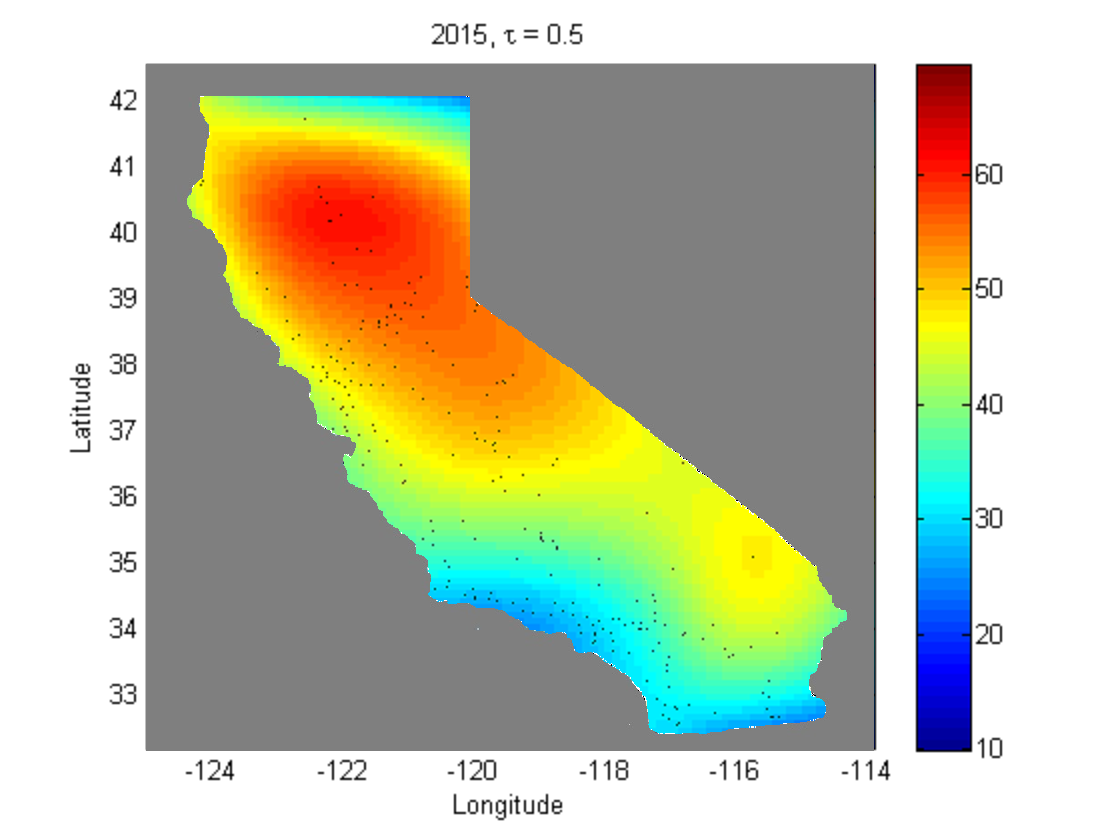} 
    \caption{Year : $2015, \tau = 0.5$}  
    \label{fig:Cali_2_50_15} 
  \end{subfigure}
    \begin{subfigure}[b]{0.5\linewidth}
    \centering
    \includegraphics[width=0.99\linewidth]{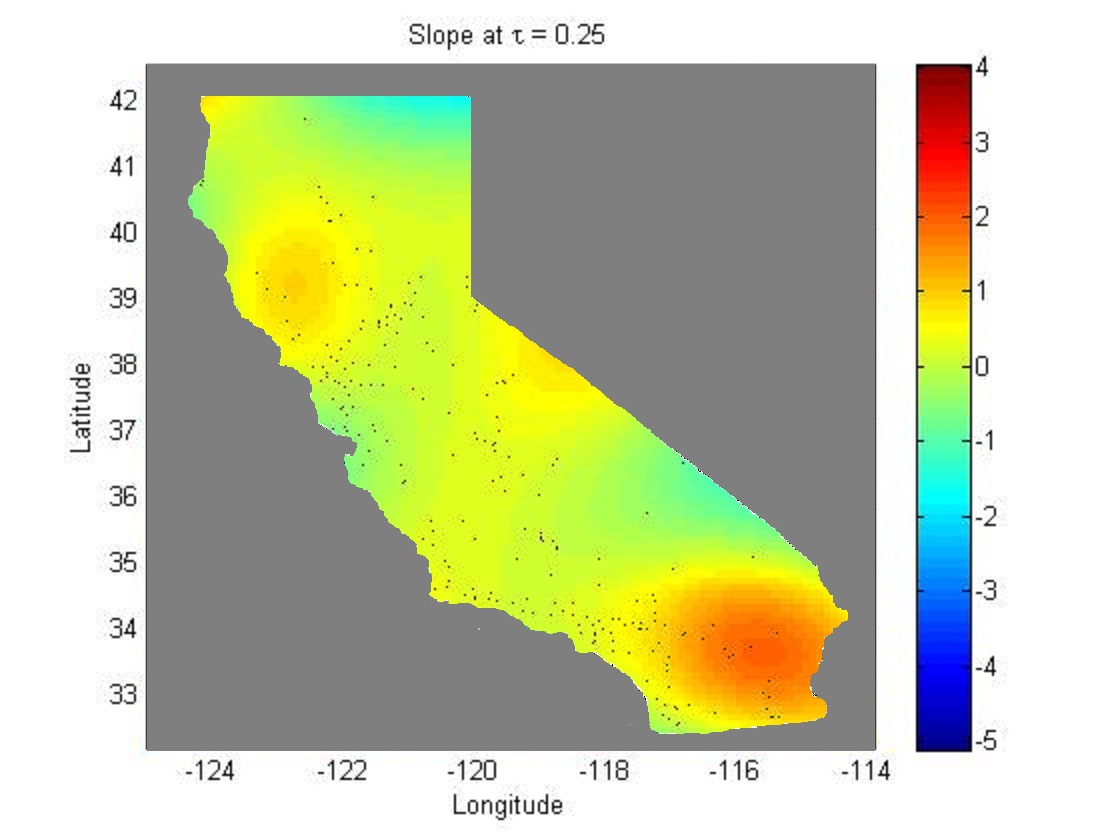} 
    \caption{Slope at $\tau = 0.25$}  
    \label{fig:CALI_SLOPE_25_8hr_BETTER} 
  \end{subfigure} \\

  \begin{subfigure}[b]{0.5\linewidth}
    \centering
    \includegraphics[width=0.99\linewidth]{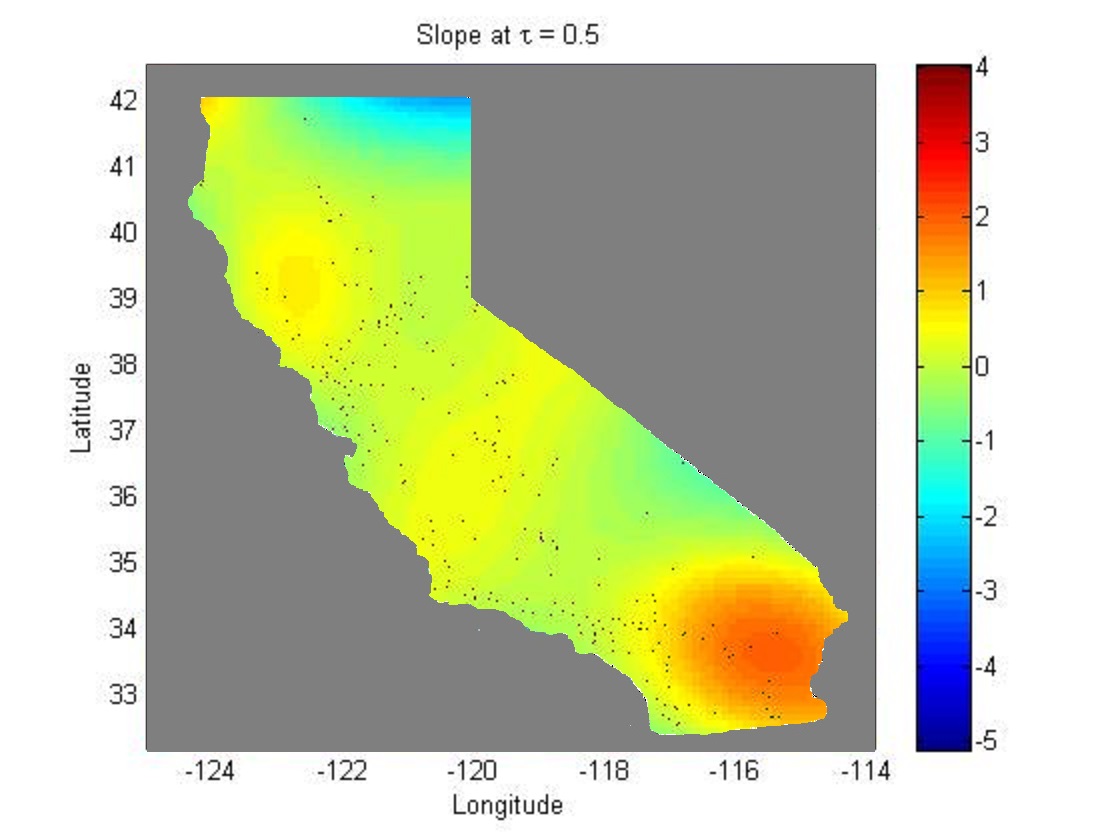} 
    \caption{Slope at $\tau = 0.5$} 
    \label{fig:CALI_SLOPE_50_8hr_BETTER} 
  \end{subfigure}
   \begin{subfigure}[b]{0.5\linewidth}
    \centering
    \includegraphics[width=0.99\linewidth]{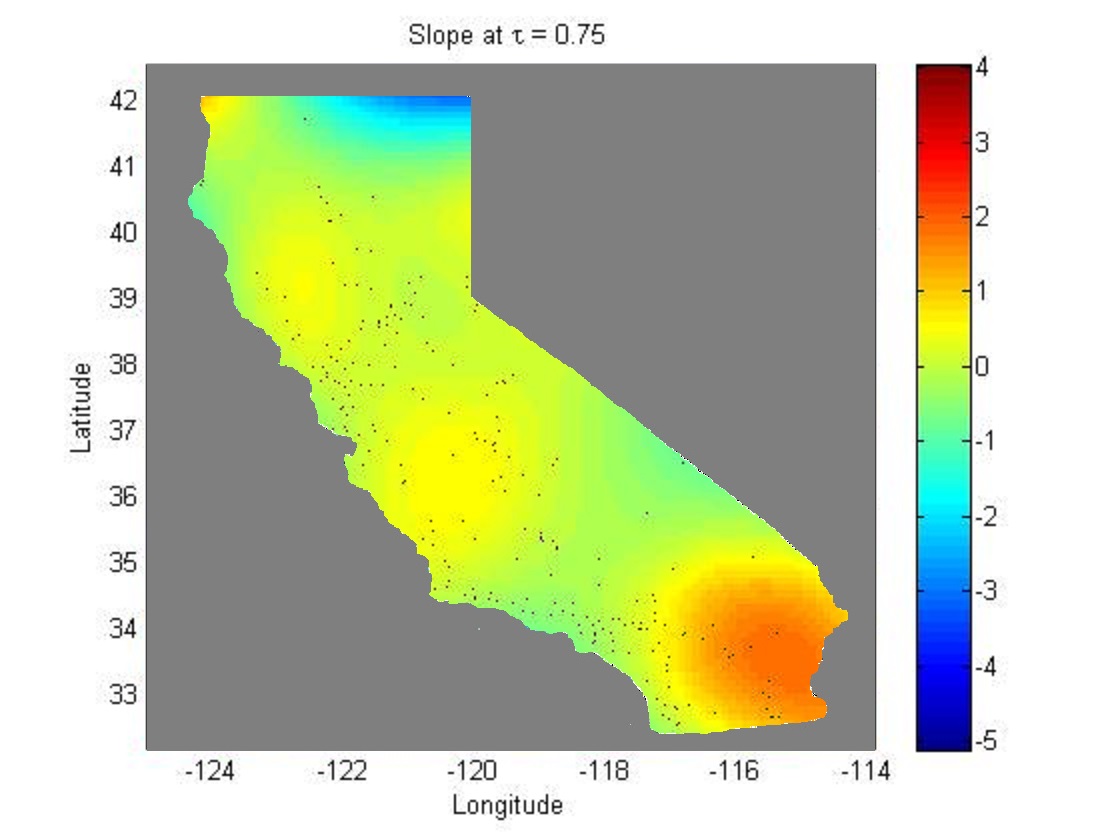} 
   \caption{Slope at $\tau = 0.75$}  
    \label{fig:CALI_SLOPE_75_8hr_BETTER} 
  \end{subfigure}
  \caption{(a-c) Daily 8-hour maximum average ozone concentration (in ppb) of California in 2006, 2010 and 2015 at $\tau = 0.5$. (d-f) Yearly rate of change of daily 8-hour maximum average ozone concentration (in ppb/year) of California at $\tau = 0.25, 0.5, 0.75$.The dots denote weather stations where data have been collected.}
  \label{figfig6} 
\end{figure}

The change in daily 8-hour maximum average ozone concentration varies a lot spatially over the time and there no particular temporal trend is observed for California as whole (see Figure \ref{figfig6}). In the High Sierra Desert and north-east part of Northern California, the 8-hour maximum average ozone concentration has a decreasing trend over time at both the quantiles. On the other hand, in Sacramento Region, Inland Empire and San Diego, it has an increasing trend over the time. Not much temporal variations have been observed in the Bay Area and the Central Coast for both 1-hour maximum and 8-hour maximum average ozone concentration levels. In High Sierra Desert and Inland Empire, this ozone concentration level is changing rapidly.  \\

\begin{figure} 
  \begin{subfigure}[b]{0.5\linewidth}
    \centering
    \includegraphics[width=0.99\linewidth]{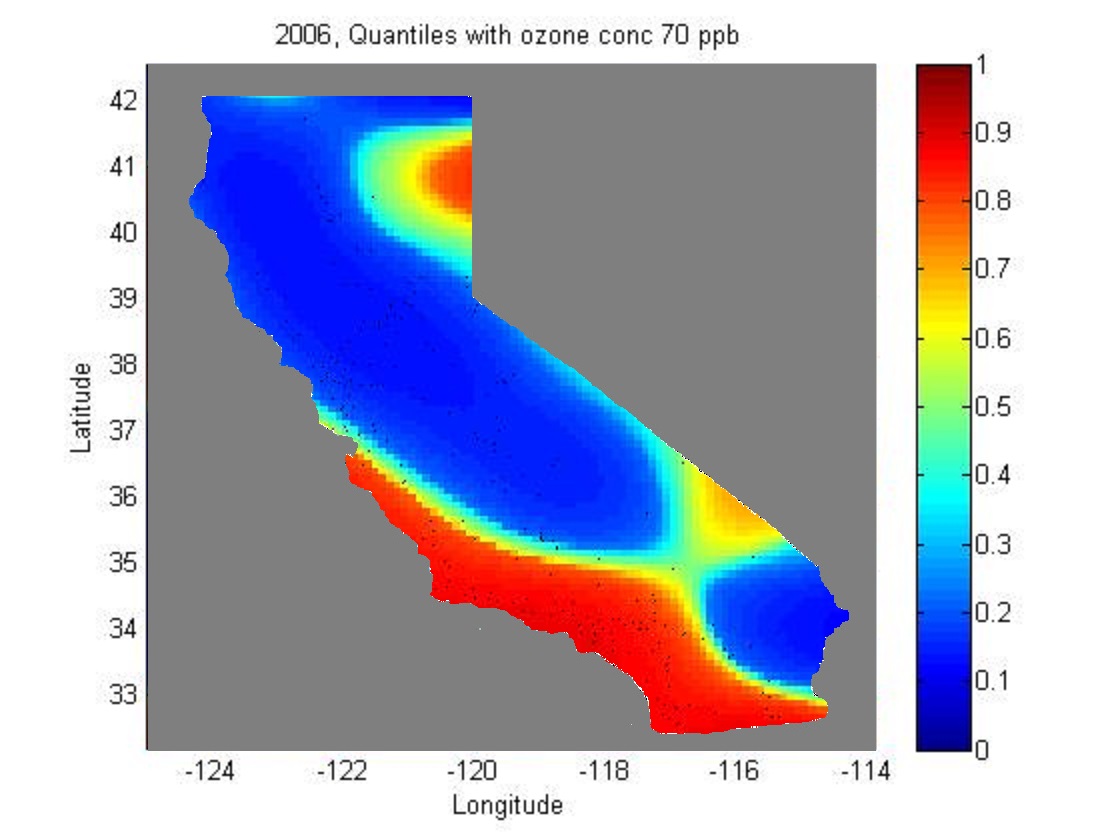} 
    \caption{2006, quantiles at 4th highest 8-hr max ozone conc. 70 ppb} 
    \label{fig:CALI_quantiles_8hr_1_better} 
  \end{subfigure}
  \begin{subfigure}[b]{0.5\linewidth}
    \centering
    \includegraphics[width=0.99\linewidth]{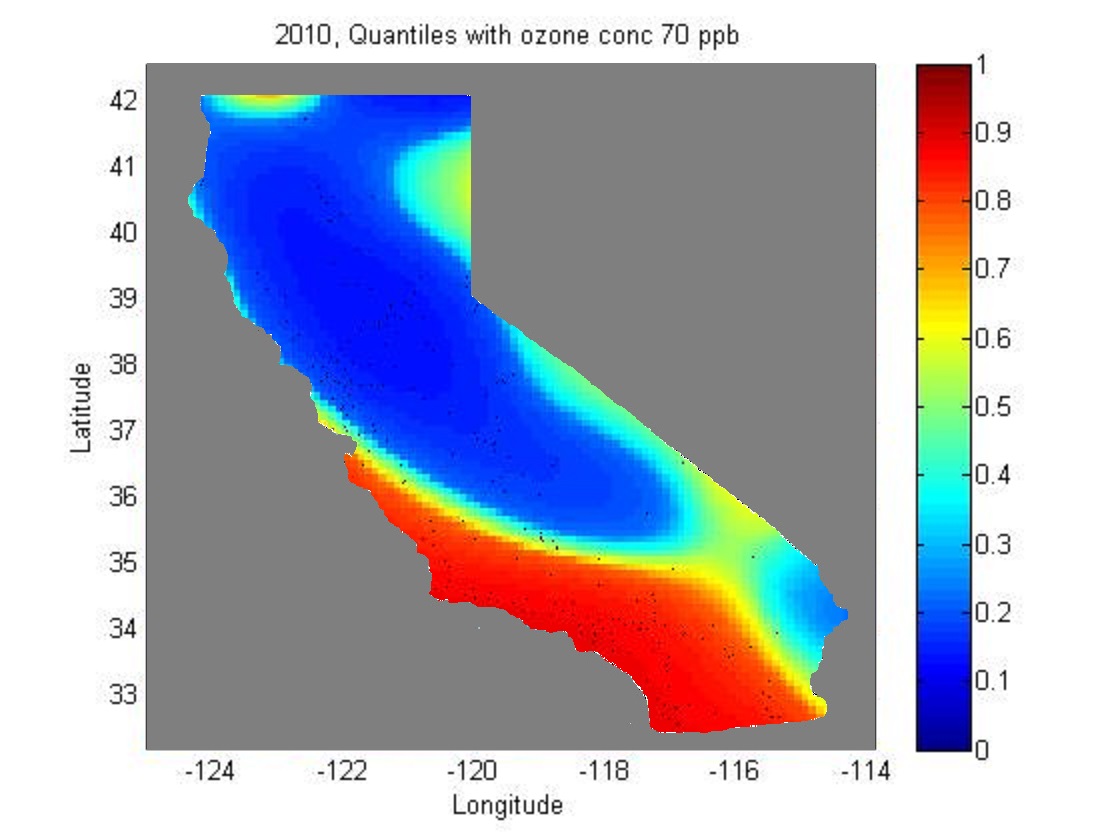} 
    \caption{2010, quantiles at 4th highest 8-hr max ozone conc. 70 ppb} 
    \label{fig:CALI_quantiles_8hr_2_better} 
  \end{subfigure}\\
   \begin{subfigure}[b]{0.5\linewidth}
    \centering
    \includegraphics[width=0.99\linewidth]{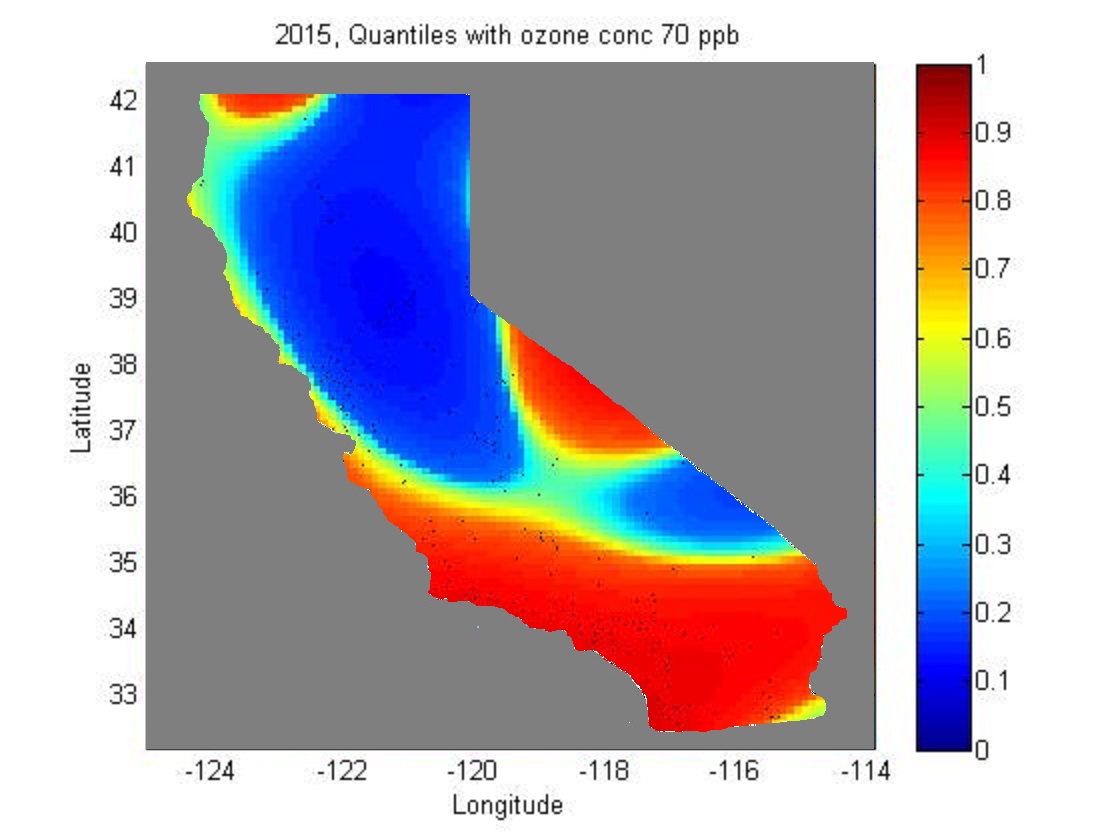} 
    \caption{2015, quantiles at 4th highest 8-hr max ozone conc. 70 ppb}  
    \label{fig:CALI_quantiles_8hr_3_better} 
  \end{subfigure}
    \begin{subfigure}[b]{0.5\linewidth}
    \centering
    \includegraphics[width=0.99\linewidth]{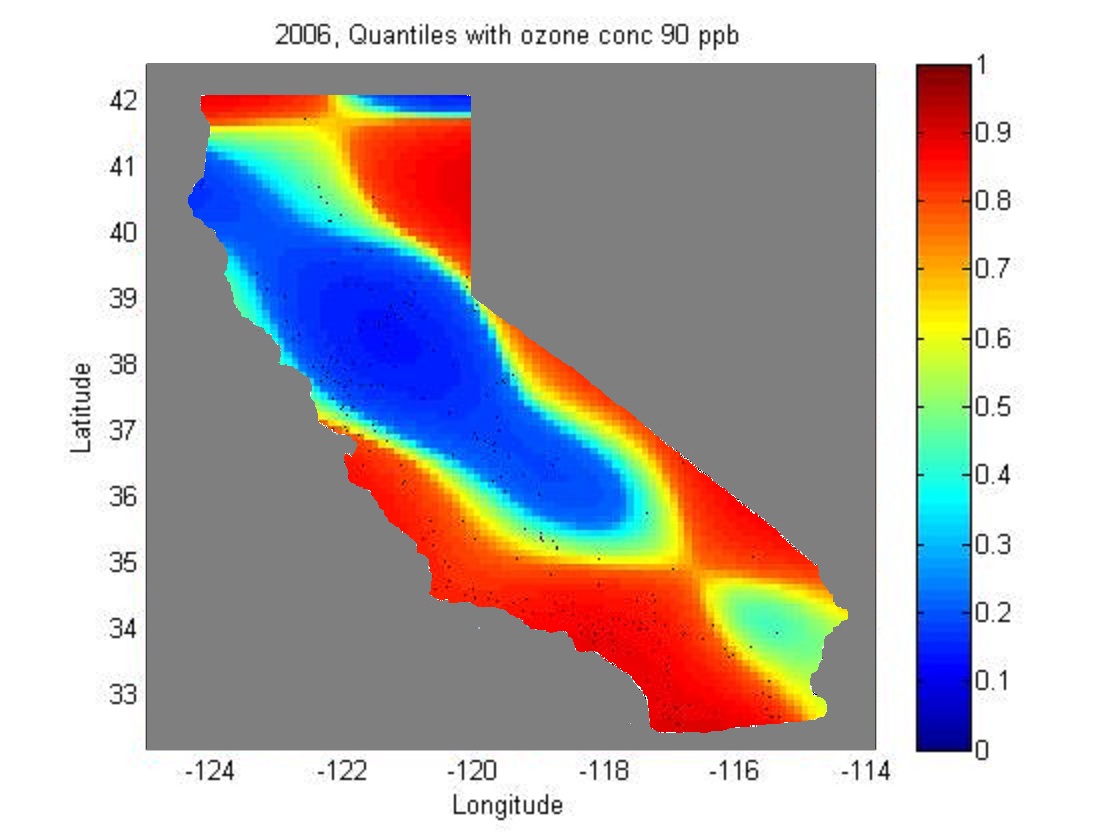} 
    \caption{2006, quantiles at 4th highest 1-hr max ozone conc. 90 ppb}  
    \label{fig:CALI_quantiles_1hr_1_better} 
  \end{subfigure} \\

  \begin{subfigure}[b]{0.5\linewidth}
    \centering
    \includegraphics[width=0.99\linewidth]{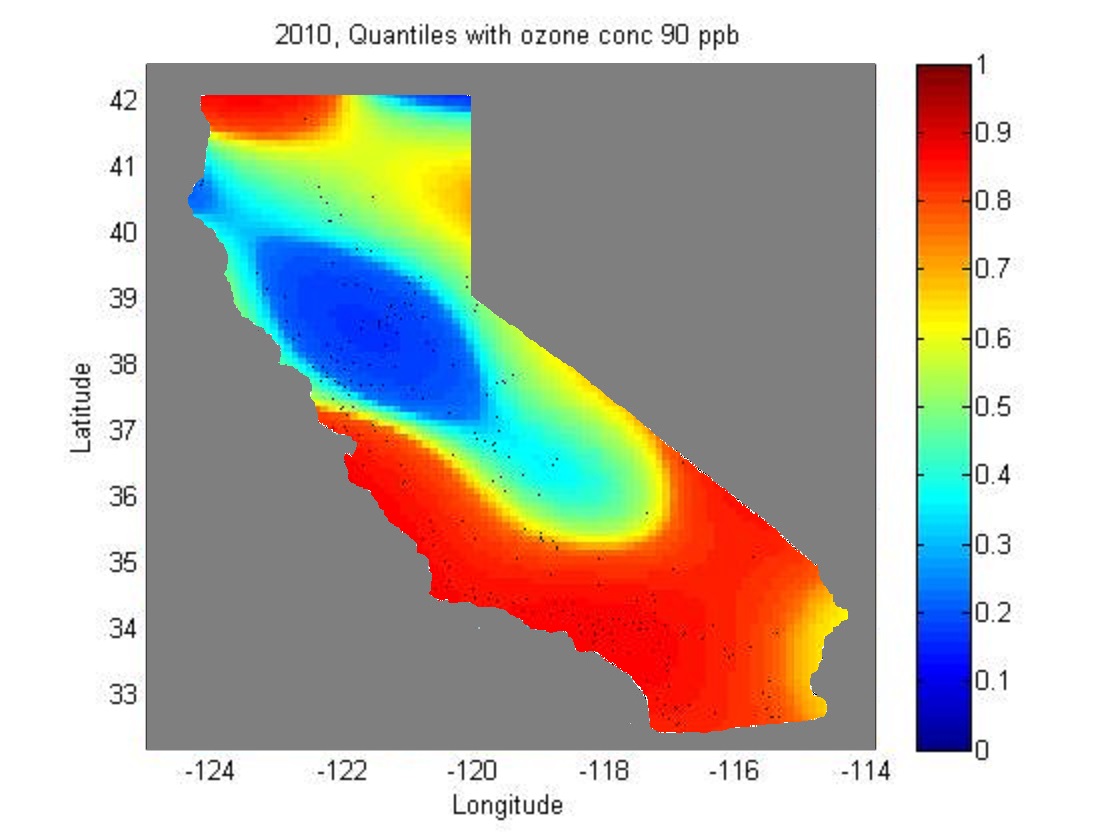} 
    \caption{2010, quantiles at 4th highest 1-hr max ozone conc. 90 ppb} 
    \label{fig:CALI_quantiles_1hr_2_better} 
  \end{subfigure}
   \begin{subfigure}[b]{0.5\linewidth}
    \centering
    \includegraphics[width=0.99\linewidth]{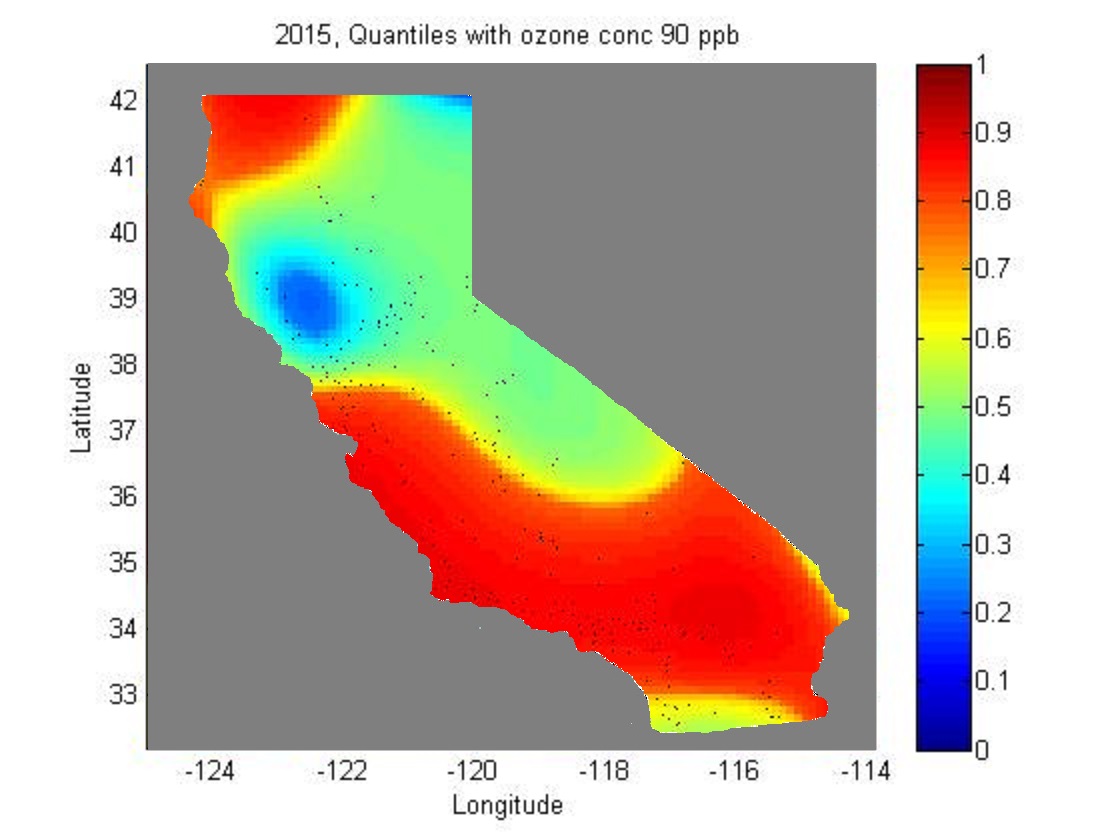} 
   \caption{2015, quantiles at 4th highest 1-hr max ozone conc. 90 ppb}  
    \label{fig:CALI_quantiles_1hr_3_better} 
  \end{subfigure}
  \caption{(a-c) Quantiles at which 4th highest daily 8-hour maximum ozone concentration (in ppb) of California is 70 ppb  in 2006, 2010 and 2015. (d-f) Quantiles at which 4th highest daily 1-hour maximum ozone concentration (in ppb) of California is 90 ppb  in 2006, 2010 and 2015.}
  \label{figfig6a} 
\end{figure}

In Figures \ref{fig:CALI_quantiles_8hr_1_better}, \ref{fig:CALI_quantiles_8hr_2_better} and \ref{fig:CALI_quantiles_8hr_3_better} it is noted that in Northern California, Bay Area, Sacramento Region and in the upper part of the Central Valley, the 4th highest daily 8-hour maximum ozone concentration threshold, i.e., 70 ppb is achieved at very lower quantile levels. In Los Angeles and San Diego area this threshold value is achieved at a higher quantile level. In Figures \ref{fig:CALI_quantiles_1hr_1_better}, \ref{fig:CALI_quantiles_1hr_2_better} and \ref{fig:CALI_quantiles_1hr_3_better} we note that at the junction of Bay Area, Sacramento Region, south Northern California and upper Central Valley the 4th highest daily 1-hour maximum ozone concentration threshold, i.e., 90 ppb is achieved at a lower quantile level compared to other parts of California. It is also noted that with time, the ozone concentration is decreasing in this region. In southern part of California, the threshold of daily 1-hour maximum ozone concentration is met at higher quantile level indicating that southern California is performing better in maintaining the ozone concentration level below recommended threshold level. But in San Diego area a strictly increasing pattern of ozone concentration is noted.

% For comaparison, we also estimated the quantile curves using the functions \textit{qreg} and \textit{qreg\_spline} of \textbf{BSquare} package in R by \citet{BSquare} which is based on the methods mentioned in \citet{Reich2012} and \citet{Reich2013}. For each of those functions, we performed 50000 iterations with 10000 burn-in. We also compared the method BPSQR (Bernstein polynomial SQR) suggested by \citet{Reich2011}, where they took Bernstein polynomial as the basis function.  \\
\section{Discussion}
As long as spatio-temporal regression is concerned, one of the major issues is the presence of missing temporal data at some locations. One way to solve this problem is to replace the missing data by estimated value based on the available data. It makes more sense to fit the model on the available data without using extrapolation or estimated missing values of data. In our regression model, the number of time-points of available data at various sites can be different (see equation (\ref{eq:likelihood})). This problem does not arise in the proposed method. \\

When the sample size is small at a location, in the method of \cite{Reich2012}, the estimated separate temporal quantile regression curve at that location will be widely affected by sampling fluctuations. If a lot of temporal data are missing at some locations, or when our data size is available only at a few temporal points, in spite of fitting separate quantile curves for all sites, it is desirable to fit a quantile curve considering the available data of all the sites together. Thus, estimation at the neighborhood of the locations with small temporal data can be improved by borrowing strength from neighboring sites. Hence the proposed method yields better estimates and is less affected by missing data and scarcity often.\\

In some approaches to spatio-temporal regression and time-series analysis, the temporal data are required to be obtained periodically. However in the proposed method, the temporal data are not required to be periodic. Thus we conclude that the proposed approach gives a natural Bayesian method for estimation and uncertainty quantification for simultaneous spatial quantile regression. The proposed method allows the data to be flexible with respect to missingness, monotonicity and low-sample size at individual locations. The method improves the quality of the inference by borrowing strength across all locations and all quantile levels by incorporating a natural non-parametric smoothing in its prior construction. By using B-spline  functions in its smoothing, the method allows a relatively more efficient approach to likelihood evaluation compared with an analogous procedure based on Gaussian procedures. \\

As mentioned in \cite{Barboza2015}, according to the EPA ozone level has decreased by about one-third over the US since 1980 by imposing regulations targeting emissions from cars, factories, consumer products and other sources of pollutants. In this analysis, it is observed that overall there is a decreasing trend in the daily 1-hour and 8-hour maximum average ozone concentration levels in the US during the period 2006-2015. Specifically, in the lest ten years, the ozone concentration has decreased considerably in the northern part of the US. In California, it is noted that there is an overall decreasing trend in the 1-hour concentration while there is no specific trend for the 8-hour concentration and the trend varies a lot spatially. According to \cite{Barboza2015}, except for a few places in California, the rest of the US is expected to comply with the EPA ozone standards by 2025. Some of the most polluted areas in Southern California and the San Joaquin Valley are expected to comply with it by 2037. 

\section*{Acknowledgements}
This project was partially supported by NSF grant number DMS 1510238.

\newpage

\appendix
%%%%

\section{Likelihood computation}
Recall, $0 = t_0 < t_1 < \cdots <t_{p_1} = 1$ are the equidistant knots on the interval $[0,1]$ such that $t_i-t_{i-1} = 1/p_1$ for $i=1,\ldots, p_1$  and $\{B_{j,m_1}(\cdot)\}_{j=1}^{p_1+m_1}$ are the basis functions of B-splines of degree $m_1$ on $[0,1]$ on the above-mentioned equidistant knots. Then Equation (\ref{eq:space_quantile}) can be written as 
\begin{align}
Q(\tau|& x,\mathbf{z}) \nonumber \\
&=  x\xi_1(\tau,\mathbf{z}) + (1-x)\xi_2(\tau,\mathbf{z}) \nonumber \\
&=  x\sum_{j=1}^{p_1+m_1}\theta_j(\mathbf{z})B_{j,m_1}(\tau)  + (1-x)\sum_{j=1}^{p_1+m_1}\phi_j(\mathbf{z})B_{j,m_1}(\tau)  \nonumber \\
&=  x \sum_{j=1}^{p_1+m_1}B_{j,m_1}(\tau)\sum_{k_1=1}^{p_2+m_2} \ldots\sum_{k_d=1}^{p_2+m_2} \alpha_{j k_1 \cdots k_d}B_{k_1,m_2}(z_1)\cdots B_{k_d,m_2}(z_d) \nonumber \\
& +(1-x)\sum_{j=1}^{p_1+m_1}B_{j,m_1}(\tau)\sum_{k_1=1}^{p_2+m_2} \ldots\sum_{k_d=1}^{p_2+m_2} \beta_{j k_1 \cdots k_d}B_{k_1,m_2}(z_1)\cdots B_{k_d,m_2}(z_d)
\end{align}
% The conditional density of $Y$ at location $\mathbf{Z} = \mathbf{z}$ is given by
% \begin{align}
% f(y|x,\mathbf{z}) & = \bigg(\frac{\partial}{\partial \tau}Q(\tau|x,\mathbf{z})\bigg|_{\tau=\tau_x(y,\mathbf{z})}\bigg)^{-1} \nonumber \\
% & = \bigg(\frac{\partial}{\partial \tau}\beta_0(\tau,\mathbf{z}) + x\frac{\partial}{\partial \tau}\beta_1(\tau,\mathbf{z})\bigg|_{\tau=\tau_x(y,\mathbf{z})}\bigg)^{-1}
% \label{eq:cond_density}
% \end{align}
% where $\tau_x(y,\mathbf{z})$ solves the equation 
% \begin{align}
% x\xi_1(\tau,\mathbf{z}) + (1-x)\xi_2(\tau, \mathbf{z}) = y.
% \label{eq:solve}
% \end{align} 

%Suppose we have data for $L$ spatial locations $\mathbf{z}_1,\ldots, \mathbf{z}_L$. Suppose $\{x_{li},y_{li}\}_{i=1}^{n_l}$ denote the values of the explanatory variable (time) and the response variable the location $\mathbf{Z} = \mathbf{z}_l$ where $n_l$ is the number of data-points at size the site $\mathbf{z}_l$ for $l=1,\ldots,L$. Then the likelihood obtained is $\prod_{l=1}^L\bigg\{\prod_{i=1}^{n_l} f(y_{li}|x_{li},\mathbf{z}_l)\bigg\}$. \\

We evaluate the likelihood by a similar approach of \cite{Tokdar2012} and \cite{Das2016}. Suppose that we have data for $L$ spatial sites $z_1,\ldots, z_L$ and at the $l$-th site, $\{X_{li},Y_{li}\}_{i=1}^{n_l}$ denote the values of the explanatory variable (time) and the response variable, $n_l$ being the number of data-points for $l=1,\ldots,L$, the likelihood is given by $\prod_{l=1}^L\bigg\{\prod_{i=1}^{n_l} f(Y_{li}|X_{li},\mathbf{z}_l)\bigg\}$ where 
\begin{align*}
f(Y_{li}|X_{li},\mathbf{z}_l) = \bigg(\frac{\partial}{\partial \tau}Q(\tau|X_{li},\mathbf{z}_l)\bigg|_{\tau=\tau_X(Y_{li},\mathbf{z}_l)}\bigg)^{-1}, \; l=1,\ldots,L, \; i = 1,\ldots,n_l.
\end{align*}
Now,
\begin{align}
\log f(Y_{li}|X_{li},\mathbf{z}_l) = & -\log\bigg(\frac{\partial}{\partial \tau}Q(\tau|X_{li},\mathbf{z}_l)\bigg|_{\tau=\tau_{X_{li}}(Y_{li},\mathbf{z}_l)}\bigg) \nonumber \\
= & -\log \bigg \{X_{li}\frac{\partial}{\partial \tau}\xi_1(\tau,\mathbf{z}_l) +  (1- X_{li})\frac{\partial}{\partial \tau}\xi_2(\tau,\mathbf{z}_l)\bigg \}\bigg|_{\tau=\tau_{X_{li}}(Y_{li},\mathbf{z}_l)}
\label{eq:single_loglike}
\end{align}
where $\tau_{X_{li}}(Y_{li},\mathbf{z}_l)$ is the solution of
\begin{align}
Y_{li} = X_{li}\xi_1(\tau,\mathbf{z}_l) + (1-X_{li})\xi_2(\tau,\mathbf{z}_l).
\label{eq:tau_sol}
\end{align}
For any given location $\mathbf{Z}=\mathbf{z}$, since $\xi_1(\cdot,\mathbf{z})$ and $\xi_2(\cdot,\mathbf{z})$ are strictly monotonic, their convex combination is also strictly monotonic. Hence there will exist a unique solution of equation (\ref{eq:tau_sol}). Using the properties of derivative of B-spline (\cite{Boor2001}) we have 
\begin{align}
\frac{d}{dt}\xi_1(t,\mathbf{z}) = \frac{d}{dt}\sum_{j=1}^{p_1+m_1}\theta_j(\mathbf{z})B_{j,m_1}(t) = \sum_{j=2}^{p_1+m_1}\theta^*_j(\mathbf{z})B_{j-1,m_1-1}(t), \nonumber \\
\frac{d}{dt}\xi_2(t,\mathbf{z}) = \frac{d}{dt}\sum_{j=1}^{p_1+m_1}\phi_j(\mathbf{z})B_{j,m_1}(t) = \sum_{j=2}^{p_1+m_1}\phi^*_j(\mathbf{z})B_{j-1,m_1-1}(t),
\label{eq:ddt}
\end{align}
where 
\begin{align*}
\theta^*_j(\mathbf{z}) = (p_1+m_1)(\theta_j(\mathbf{z})-\theta_{j-1}(\mathbf{z})), \;  \phi^*_j(\mathbf{z}) = (p_1+m_1)(\phi_j(\mathbf{z})-\phi_{j-1}(\mathbf{z})),
\end{align*}
for $j = 2,\ldots,p_1+m_1$.
Now, using equation (\ref{eq:single_loglike}) and (\ref{eq:ddt}), we have
\begin{align}
\log f(Y_{li}|X_{li},\mathbf{z}_l) = & -\log \bigg\{ X_{li} \sum_{j=2}^{p_1+m_1}\theta^*_j(\mathbf{z}_l)B_{j-1,m_1-1}(\tau_{X_{li}}(Y_{li},\mathbf{z}_l))  \nonumber \\
& + (1-X_{li})\sum_{j=2}^{p_1+m_1}\phi^*_j(\mathbf{z}_l)B_{j-1,m_1-1}(\tau_{X_{li}}(Y_{li},\mathbf{z}_l)) \bigg\}.
\end{align}
Hence the total log-likelihood is given by
\begin{align}
\sum_{l=1}^L\sum_{i=1}^{n_l}\log f(Y_{li}|X_{li},\mathbf{z}_l) = & -\sum_{l=1}^L\sum_{i=1}^{n_l}\log \bigg\{ X_{li} \sum_{j=2}^{p_1+m_1}\theta^*_j(\mathbf{z}_l)B_{j-1,m_1-1}(\tau_{X_{li}}(Y_{li},\mathbf{z}_l))   \nonumber \\
& + (1-X_{li})\sum_{j=2}^{p_1+m_1}\phi^*_j(\mathbf{z}_l)B_{j-1,m_1-1}(\tau_{X_{li}}(Y_{li},\mathbf{z}_l)) \bigg\}.
\end{align}
We note that the parameters of the likelihood are the coefficients of the B-spline basis expansion of $\{\theta_j(\mathbf{z})\}_{j=1}^{p_1+m_1}$ and $\{\phi_j(\mathbf{z})\}_{j=1}^{p_1+m_1}$ which are 
\begin{align}
0 = \alpha_{1k_1\cdots k_d} < \cdots < \alpha_{(p_1+m_1)k_1\cdots k_d} = 1,  \nonumber \\
0 = \beta_{1k_1\cdots k_d} < \cdots < \beta_{(p_1+m_1)k_1\cdots k_d} = 1,
\label{eq:parameters}
\end{align}
for $\{k_1,\ldots, k_d\} \in \{1,\ldots, (p_2+m_2)\}^d$.

\section{Block Metropolis-Hastings MCMC algorithm}
Recall that
\begin{align}
\gamma_{jk_1\cdots k_d} = \alpha_{(j+1)k_1\cdots k_d} - \alpha_{jk_1\cdots k_d}, \; 
\delta_{jk_1\cdots k_d} = \beta_{(j+1)k_1\cdots k_d} - \beta_{jk_1\cdots k_d},
\end{align}
for $j=1,\ldots,p_1+m_1-1$ and $\{k_1,\ldots, k_d\} \in \{1,\ldots, (p_2+m_2)\}^d$. It is noted that $\{\gamma_{jk_1\cdots k_d}\}_{j=1}^{p_1+m_1-1}$ and $\{\delta_{jk_1\cdots k_d}\}_{j=1}^{p_1+m_1-1}$ are on the unit simplex for any given $\{k_1,\ldots, k_d\} \in \{1,\ldots, (p_2+m_2)\}^d$. We use Block Metropolis-Hastings Monte Carlo Markov Chain (MCMC) algorithm (see \cite{Chib1995}) for sampling from the posterior distribution. Note that, the number of unit simplex blocks is $(p_2+m_2)^d$. In MCMC, a move is initiated on each unit simplex block in a loop. During the updating stage of a single unit-simplex block, similar to the updating strategy used in \cite{Das2016}, independent sequences $\{U_j\}_{j=1}^{p_1+m_1-1}$ and $\{W_j\}_{j=1}^{p_1+m_1-1}$ are generated from $U(1/r,r)$ for some $r>1$. For given $\{k_1,\ldots, k_d\} \in \{1,\ldots, (p_2+m_2)\}^d$, define $V_j=\gamma_{jk_1\cdots k_d} U_j$ and $T_j=\delta_{jk_1\cdots k_d} W_j$ for $j=1,\ldots,p_1+m_1-1$. The proposal moves $\gamma_{jk_1\cdots k_d} \mapsto \gamma_{jk_1\cdots k_d}^*$ and $\delta_{jk_1\cdots k_d} \mapsto \delta_{jk_1\cdots k_d}^*$ are given by 
\begin{align}
\gamma_{jk_1\cdots k_d}^* = \frac{V_j}{\sum_{i=1}^{p_1+m_1-1}V_i}, \; \delta_{jk_1\cdots k_d}^* = \frac{T_j}{\sum_{i=1}^{p_1+m_1-1}T_i}, \; j=1,\ldots,p_1+m_1-1. 
\end{align}
The conditional distribution of $\{\gamma_{jk_1\cdots k_d}^*\}_{j=1}^{p_1+m_1-1}$ given $\{\gamma_{jk_1\cdots k_d}\}_{j=1}^{p_1+m_1-1}$ is given by (see appendix of \cite{Das2016} for the derivation)
\begin{align}
f(\mathbf{\gamma}_{.k_1\cdots k_d}^*|\mathbf{\gamma}_{.k_1\cdots k_d})= \bigg(\frac{r}{r^2-1}\bigg)^{p_1+m_1-1}\bigg\{\prod\limits_{j=1}^{p_1+m_1-1} \gamma_{jk_1\cdots k_d}\bigg\}^{-1}\frac{(D_1-D_2)}{(p_1+m_1-1)},
\end{align}
where 
\begin{align*}
D_1 = & \Big(\min_{0 \leq j \leq p_1+m_1-1}\ \frac{r\gamma_{jk_1\cdots k_d}}{\gamma_{jk_1\cdots k_d}^*}\Big)^{p_1+m_1-1}, \\ 
D_2 = & \Big(\max_{0 \leq j \leq p_1+m_1-1} \frac{\gamma_{jk_1\cdots k_d}}{r\gamma_{jk_1\cdots k_d}^*}\Big)^{p_1+m_1-1}.
\end{align*}
The conditional distribution of $\{\delta_{jk_1\cdots k_d}^*\}_{j=1}^{p_1+m_1-1}$ can be found in a similar way. The updated values of $\{\alpha_{jk_1\cdots k_d}\}_{j=1}^{p_1+m_1-1}$,$\{\beta_{jk_1\cdots k_d}\}_{j=1}^{p_1+m_1-1}$, denoted by $\{\alpha^*_{jk_1\cdots k_d}\}_{j=1}^{p_1+m_1-1}$,$\{\beta^*_{jk_1\cdots k_d}\}_{j=1}^{p_1+m_1-1}$, can be found from $\{\gamma^*_{jk_1\cdots k_d}\}_{j=1}^{p_1+m_1-1}$ and $\{\delta^*_{jk_1\cdots k_d}\}_{j=1}^{p_1+m_1-1}$ using the relation
\begin{align*}
\alpha_{jk_1\cdots k_d}^*=\sum_{i=1}^{j}\gamma_{ik_1\cdots k_d}^*, \; \beta_{jk_1\cdots k_d}^*=\sum_{i=1}^{j}\delta_{ik_1\cdots k_d}^*, \quad j=1, \ldots, p_1+m_1-1. \\
\label{eq : relation_4}
\end{align*}
The likelihood can be expressed as the function of $\{\alpha_{jk_1\cdots k_d},\beta_{jk_1\cdots k_d}\}_{j=1}^{p_1+m_1-1}$ assuming the values of the other unit simplex blocks to be fixed. Hence, it can be also expressed as a function of $\{\gamma_{jk_1\cdots k_d},\delta_{jk_1\cdots k_d}\}_{j=1}^{p_1+m_1-1}$. Let $L(\gamma_{.k_1\cdots k_d},\delta_{.k_1\cdots k_d})$ and $L(\gamma^*_{.k_1\cdots k_d},\delta^*_{.k_1\cdots k_d})$ denote the likelihood at $\{\gamma_{jk_1\cdots k_d},$ $\delta_{jk_1\cdots k_d}\}_{j=1}^{p_1+m_1-1}$ and $\{\gamma^*_{jk_1\cdots k_d},\delta^*_{jk_1\cdots k_d}\}_{j=1}^{p_1+m_1-1}$ respectively fixing the values of the parameters of the other unit simplex blocks. The acceptance probability in the Block Metropolis-Hastings algorithm for the update step of the corresponding block is given by $P_a= \min \{p, 1\}$ where 
\begin{align*}
p = & \frac{L(\gamma^*_{.k_1\cdots k_d},\delta^*_{.k_1\cdots k_d})\pi(\gamma^*_{.k_1\cdots k_d})\pi(\delta^*_{.k_1\cdots k_d})f(\mathbf{\gamma}_{.k_1\cdots k_d}|\mathbf{\gamma}^*_{.k_1\cdots k_d})f(\mathbf{\delta}_{.k_1\cdots k_d}|\mathbf{\delta}^*_{.k_1\cdots k_d})}{L(\gamma_{.k_1\cdots k_d},\delta_{.k_1\cdots k_d})\pi(\gamma_{.k_1\cdots k_d})\pi(\delta_{.k_1\cdots k_d})f(\mathbf{\gamma}_{.k_1\cdots k_d}^*|\mathbf{\gamma}_{.k_1\cdots k_d})f(\mathbf{\delta}_{.k_1\cdots k_d}^*|\mathbf{\delta}_{.k_1\cdots k_d})} \\
= & \frac{L(\gamma^*_{.k_1\cdots k_d},\delta^*_{.k_1\cdots k_d})f(\mathbf{\gamma}_{.k_1\cdots k_d}|\mathbf{\gamma}^*_{.k_1\cdots k_d})f(\mathbf{\delta}_{.k_1\cdots k_d}|\mathbf{\delta}^*_{.k_1\cdots k_d})}{L(\gamma_{.k_1\cdots k_d},\delta_{.k_1\cdots k_d})f(\mathbf{\gamma}_{.k_1\cdots k_d}^*|\mathbf{\gamma}_{.k_1\cdots k_d})f(\mathbf{\delta}_{.k_1\cdots k_d}^*|\mathbf{\delta}_{.k_1\cdots k_d})}
\end{align*}
and $\pi$ denotes the prior density. Since we take uniform Dirichlet prior on the unit simplex blocks $\{\gamma_{jk_1\cdots k_d}\}_{j=1}^{p_1+m_1-1}$ and $\{\delta_{jk_1\cdots k_d}\}_{j=1}^{p_1+m_1-1}$, each unit simplex block is updating one at a time in a loop.

\label{lastpage}


\begin{thebibliography}{}

\bibitem[Barboza (2015)]{Barboza2015} Barboza T. 2015. New attack on California's dirty air, {\it Los Angeles Times}, Available at  http://www.latimes.com/local/lanow/la-me-ln-what-new-smog-rules-mean-for-california-air-pollution-20150930-story.html.

\bibitem[Bethke (1980)]{Bethke1980}  Bethke A. 1980. Genetic algorithms as function optimizers, Available at  https://deepblue.lib.umich.edu/handle/2027.42/3572.

\bibitem[Bondell et~al.(2010)]{Bondell2010} Bondell HD, Reich RJ, Wang H. 2010. Non-crossing quantile regression curve estimation, {\it Biometrika}, {\bf 97}:  825--838.


\bibitem[Chib and Greenberg (1995)]{Chib1995}  Chib S, Greenberg E. 1995. Understanding the Metropolis-Hastings algorithm, {\it The American Statistician}, {\bf 49(4)}: 327--335.

\bibitem[Das (2016a)]{DAS_3_2016} Das P. 2016. Black-box optimization on multiple simplex constrained blocks, {\it arXiv:1609.02249v1}.

\bibitem[Das (2016b)]{DAS_1_2016} Das P. 2016. Derivative-free efficient global optimization of function of parameters from bounded intervals, {\it arXiv:1604.08616v1}.

\bibitem[Das (2016c)]{DAS_2_2016} Das P. 2016. Derivative-free efficient global optimization on high-dimensional simplex, {\it arXiv:1604.08636v1}.

\bibitem[Das and Ghosal (2016)]{Das2016} Das P, Ghosal S. 2016. Bayesian quantile regression using random B-spline series prior,{\it Computational Statistics and Data Analysis (To appear, DOI : 10.1016/j.csda.2016.11.014) arXiv:1609.02950v1}.

\bibitem[Boor (2001)]{Boor2001} Boor C. 2001. {\it A Practical Guide to Splines}, Revised Edition, New-York : Springer.

\bibitem[Dunson and Taylor (2005)]{Dunson2005} Dunson DB, Taylor J. 2005. Approximate Bayesian inference for quantiles, {\it Journal of Nonparametric Statistics}, {\bf 17(3)}: 385--400.

\bibitem[Dunson and Park (2008)]{Dunson2008} Dunson DB, Park J. 2008. Kernel stick-breaking processes, {\it Biometrika}, {\bf 95}: 307--323.

\bibitem[Fraser (1957)]{Fraser1957} Fraser A. 1957. Simulation of genetic systems by automatic digital computers, {\it Australian Journal of Biological Sciences}, {\bf 10}: 484--491.

\bibitem[Gelfand and Kottas (2003)]{Gelfand2003} Gelfand AE, Kottas A. 2003. Bayesian semiparametric regression model for median residual life, {\it Scandinavian Journal of Statistics}, {\bf 30(4)}: 651--665.

\bibitem[Gelfand et~al.(2005)]{Gelfand2005} Gelfand AE, Kottas A, MacEachern SN. 2005. Bayesian non-parametric spatial modeling with Dirichlet process mixing, {\it Journal of the American Statistical Association}, {\bf 100}: 1021--1035.

\bibitem[Goldberg (1989)]{Goldberg1989} Goldberg D. 1989. {\it Genetic Algorithms in Search, Optimization and Machine Learning}, Addison-Wesley Publishing Company.

\bibitem[Granville et~al.(1994)]{Granville1994} Granville V, Krivanek M, Rasson J. 1994. Simulated annealing: A proof of convergence, {\it IEEE Transactions on Pattern Analysis and Machine Intelligence}, {\bf 16}: 652--656.

\bibitem[Griffin and Steel (2006)]{Griffin2006} Griffin JE, Steel MFJ. 1994. Order-based dependent Dirichlet processes, {\it Journal of the American Statistical Association}, {\bf 101}: 179--194.

\bibitem[He (1997)]{He1997} He X. 1997. Quantile curves without crossing, {\it American Statistician}, {\bf 51}: 186--192.

\bibitem[Kirkpatrick et~al.(1983)]{Kirkpatrick1983} Kirkpatrick S,  Gelatt Jr C, Vecchi M. 1983. Optimization by Simulated Annealing, {\it Australian Journal of Biological Sciences}, {\bf 220}: 671--680.

\bibitem[Koenkar (2005)]{Koenkar2005} Koenkar R. 2005. {\it Quantile Regression}, London: Cambridge University Press.

\bibitem[Koenkar and Bassett (1978)]{Koenkar1978} Koenkar R, Bassett G. 1978. Regression quantiles, {\it Econometrica}, {\bf 46}: 33--50.

\bibitem[Kottas and Gelfand (2001)]{Kottas2001} Kottas A, Gelfand A. 2001. Bayesian semiparametric median regression modeling, {\it Journal of the American Statistical Association}, {\bf 96}: 1458--1468.

\bibitem[Kottas and Krnjajic (2009)]{Kottas2009} Kottas A, Krnjajic M. 2009. Bayesian semiparametric modelling in quantile regression, {\it Scandinavian Journal of Statistics}, {\bf 36}: 297--319.

\bibitem[Lee and Neocleous (2010)]{Lee2010} Lee D, Neocleous T. 2010. Bayesian quantile regression for count data with application to environmental epidemiology, {\it Journal of the Royal Statistical Society: Series C}, {\bf 59}: 905--920.

\bibitem[Neocleous and Portnoy (2008)]{Neocleous2008} Neocleous T, Portnoy S. 2008. Quantile curves without crossing, {\it Statistics and Probability Letters}, {\it 78}: 1226--1229.

\bibitem[Oh et~al.(2011)]{Oh2011} Oh H, Lee T, Nychka D. 2011. Fast non-parametric quantile regression with arbitrary smoothing methods, {\it Journal of Computational and Graphical Statistics}, {\bf 20}: 510--526.

\bibitem[Reich (2012)]{Reich2012} Reich BJ. 2012. Spatiotemporal quantile regression for detecting distributional changes in environmental processes, {\it Journal of Royal Statistical Society Series C (Applied Statistics)}, {\bf 61(4)}: 535--553.

\bibitem[Reich and Fuentes (2007)]{Reich2007} Reich BJ, Fuentes M. 2007. A multivariate semiparametric Bayesian spatial modeling framework for hurricane surface wind fields, {\it the Annals of Applied Statistics}, {\bf 1}: 249--264.

\bibitem[Reich et~al.(2011)]{Reich2011} Reich B, Fuentes M, Dunson D. 2011. Bayesian spatial quantile regression, {\it Journal of the American Statistical Association}, {\bf 106(493)}: 6--20.

\bibitem[Shim et~al.(2009)]{Shim2009} Shim J, Hwang C, Seok K. 2009.  Non-crossing quantile regression via doubly penalized kernel machine, {\it Computational Statistics}, {\bf 24}: 83--94.

\bibitem[Shim and Lee (2010)]{Shim2010} Shim J, Lee J. 2010. Restricted support vector quantile regression without crossing, {\it Journal of the Korean Data and Information Science Society}, {\bf 21(6)}: 1319--1325.

\bibitem[Sobotka and Kneib (2012)]{Sobotka2012} Sobotka F, Kneib T. 2012. Geoadditive expectile regression, {\it Computational Statistics and Data Analysis}, {\bf 56(4)}: 755--767.

\bibitem[Takeuchi (2004)]{Takeuchi2004} Takeuchi I. 2004. Non-crossing quantile regression curves by support vector and its effcient implementation, {\it Proceedings of 2004 IEEE IJCNN}, {\bf 1}: 401--406.

\bibitem[Takeuchi et~al.(2006)]{Takeuchi2006} Takeuchi T, Le Q, Sears T, Smola AJ. 2006. Noparametric quantile estimation, {\it Journal of Machine Learning Research}, {\bf 7}: 1231--1264.

\bibitem[Tokdar and Kadane (2012)]{Tokdar2012} Tokdar S, Kadane JB. 2012. Simultaneous linear quantile regression : a semiparametric Bayesian approach, {\it Bayesian Analysis}, {\bf 7(1)}: 51--72.


\bibitem[Vapnik (1995)]{Vapnik1995} Vapnik V. 1995. {\it The Nature of Statistical Learning Theory}, New York: Springler-Verlag.

\bibitem[Wu and Liu (2009)]{Wu2009} Wu Y, Liu Y. 2009. Stepwise multiple quantile regression estimation using non-crossing constraints, {\it Statistics and Its Inference}, {\bf 2}: 299--310.

\bibitem[Yu and Moyeed (2001)]{Yu2001} Yu K, Moyeed RA. 2001. Bayesian quantile regression, {\it Statistics and Probability Letters}, {\bf 54}: 437--447.

\end{thebibliography}
\end{document}